\documentclass[11pt,a4paper]{article}

\usepackage[a4paper,margin=1in]{geometry}
\usepackage[T1]{fontenc}
\usepackage[utf8]{inputenc}
\usepackage{lmodern}
\usepackage{amsmath,amssymb,mathtools,bm,mathrsfs}
\usepackage{physics}
\usepackage{microtype}
\usepackage{hyperref}
\usepackage{booktabs}
\usepackage{cite}
\usepackage{authblk}
\usepackage{setspace}
\usepackage{amsthm}

\numberwithin{equation}{section}
\newtheorem{remark}{Remark}
\hypersetup{colorlinks=true,citecolor=blue,linkcolor=blue,urlcolor=blue}

\newcommand{\R}{\mathbb{R}}
\newcommand{\F}{\mathcal{F}}
\newcommand{\Hh}{\mathcal{H}}
\newcommand{\D}{\mathscr{D}}
\newcommand{\ep}{\epsilon}
\newcommand{\ii}{\mathrm{i}}
\newcommand{\ee}{\mathrm{e}}
\newcommand{\w}{\omega}

\begin{document}

\title{Renormalization of One-Dimensional Semirelativistic Bosons
with Contact Interactions}

\author[1]{Fatih Erman}
\author[2]{O. Teoman Turgut}
\affil[1]{Department of Mathematics, \.{I}zmir Institute of Technology, Urla, 35430, \.{I}zmir, Turkey}
\affil[2]{Department of Physics, Bo\u{g}azi\c{c}i University, Bebek, 34342, \.{I}stanbul, Turkey}
\affil[1]{fatih.erman@gmail.com}
\affil[2]{turgutte@boun.edu.tr}

\date{\today}
\maketitle

\begin{abstract}
We study one-dimensional spinless bosons with semirelativistic
spinless-Salpeter dispersion and attractive pairwise contact interactions.
Because the dispersion becomes linear at large momentum, the contact
interaction is marginal by power counting and produces a logarithmic
ultraviolet divergence.  We construct the renormalized many-body theory
using an enlarged Fock space and a Schur-complement representation of the
resolvent, eliminating the bare coupling in favor of the physical
zero-total-momentum two-body bound-state energy.  The resulting
cutoff-independent resolvent defines a self-adjoint Hamiltonian in each
fixed particle-number sector.  We treat the two-body problem explicitly
and show, in the norm-resolvent sense, that the nonrelativistic limit
reproduces the attractive Lieb--Liniger Hamiltonian.  We also formulate a
mean-field approximation directly within the renormalized theory.  In the
massless and deeply bound large-particle-number regimes, it predicts an
exponentially increasing binding scale whose exponent is determined by a
one-dimensional variational problem.
\end{abstract}

\section{Introduction}
\label{sec:intro}

Contact interactions provide some of the simplest settings in which the
interplay between few-body physics, many-body correlations, and
renormalization can be studied explicitly.  In one spatial dimension the
standard nonrelativistic example is the Lieb--Liniger model.  In its
original finite-volume formulation it describes bosons on a ring, or
equivalently on an interval with periodic boundary conditions, and admits
a Bethe-ansatz solution \cite{LiebLiniger1963}.  The attractive model
supports many-particle bound states
\cite{McGuire1964,Yang1967,Yang1968} and also provides a useful setting
in which exact many-body results can be compared with Hartree and
mean-field approximations \cite{CalogeroDegasperis1975}.  An important
feature of the one-dimensional nonrelativistic contact problem is that
the two-body delta interaction does not produce an ultraviolet divergence,
so no coupling-constant renormalization is required.

We replace the quadratic kinetic energy by the spinless-Salpeter
dispersion and consider identical spinless bosons on the full line
$\mathbb R$.  Formally, in the $n$-particle sector the model is
\begin{equation}
 H_{n,\mathrm{formal}}
 =
 \sum_{\alpha=1}^{n}
 \sqrt{\hat p_\alpha^{\,2}+m^2}
 -
 \lambda\sum_{\alpha<\beta}
 \delta(x_\alpha-x_\beta),
 \qquad \lambda>0,
 \label{eq:formalhamiltonian}
\end{equation}
where $\hbar=c=1$ unless stated otherwise.  We shall refer to this system
as a semirelativistic analogue of the attractive Lieb--Liniger model.
The terminology refers only to replacing the nonrelativistic
one-particle dispersion while retaining the same pairwise contact
structure.  The interaction is instantaneous and particle number is
conserved; the model is therefore not a Lorentz-invariant quantum field
theory and should not be confused with relativistic integrable field
theories whose nonrelativistic limits are related to the Lieb--Liniger
model \cite{Kormos2009}.

The need for renormalization can already be anticipated from ultraviolet
scaling.  At large momentum,
\begin{equation}
 \sqrt{p^2+m^2}\sim |p|,
 \qquad |p|\to\infty.
\end{equation}
Under a spatial dilation $x\mapsto a^{-1}x$, one has $p\mapsto ap$.
The ultraviolet kinetic energy therefore scales as
$|\hat p|\mapsto a|\hat p|$, while in one dimension
$\delta(x)\mapsto a\,\delta(x)$.  Thus the kinetic and contact terms have
the same ultraviolet scaling dimension and the coupling is dimensionless
by power counting.  The contact interaction is consequently marginal.
This is in sharp contrast with the ordinary nonrelativistic dispersion,
for which $p^2$ and $\delta(x)$ scale differently.

The logarithmic behavior predicted by power counting is visible already
in the two-body sector.  At zero total momentum, free propagation of a
pair between two contact configurations contains
\begin{equation}
 \int_{-\infty}^{\infty}\frac{\dd k}{2\pi}\,
 \frac{1}{2\sqrt{k^2+m^2}-E}.
 \label{eq:introcontactintegral}
\end{equation}
The integrand behaves as $1/(2|k|)$ for large $|k|$, and the integral
therefore diverges logarithmically.  Equivalently, the short-time
behavior of the Salpeter heat kernel gives
\begin{equation}
 K_{2t}(0)\sim\frac{1}{2\pi t},
 \qquad t\to0^+,
\end{equation}
which produces the same logarithm after integration over short times.
The one-dimensional Salpeter problem with Dirac delta interactions is
already known to require coupling-constant renormalization
\cite{AlHashimi2014,Albeverio2015,ErmanGadellaUncu2017}; the present
problem asks how this ultraviolet structure extends to an interacting
many-boson system.

To construct the theory we use the enlarged-Fock-space method originally
introduced by S.~G. Rajeev for singular point interactions
\cite{Rajeev1999} and subsequently developed for related many-body
systems \cite{ErmanTurgut2010,DoganErmanTurgut2012,ErmanTurgut2013}.
The bosonic Fock space is enlarged by adjoining a continuous
orthofermionic algebra.  The auxiliary degree of freedom does not
represent an additional physical particle; rather, it records the
contact channel associated with a pair of bosons.  The regularized
problem then acquires a $2\times2$ block structure with respect to the
zero- and one-orthofermion sectors.

The two complementary Schur complements have direct meanings.
Eliminating the one-orthofermion sector reproduces the regularized
bosonic Hamiltonian, whereas eliminating the bosonic sector yields an
energy-dependent operator $\Phi_\epsilon(E)$ acting in the auxiliary
contact-pair sector.  Following the terminology of this approach, we call
this reduced operator the \emph{principal operator}.  Its normal-ordered
form separates the primitive logarithmically divergent two-body bubble
from terms that remain finite when the regulator is removed.  The bare
coupling is eliminated in favor of the physical zero-total-momentum
two-body bound-state energy $E_B<2m$.  In the massless limit, replacing a
dimensionless bare coupling by a physical energy scale is the familiar
mechanism of dimensional transmutation.

After renormalization, the bosonic resolvent takes the form
\begin{equation}
 R(E)
 =
 R_0(E)
 +
 T(\overline E)^\dagger
 \Phi(E)^{-1}
 T(E),
 \label{eq:intro-ren-res}
\end{equation}
where
\begin{equation}
 T(E)
 =
 \lim_{\epsilon\to0^+}
 b_\epsilon R_0(E)
\end{equation}
is the resolvent-dressed contact map.  This formulation avoids treating
the unsmeared zero-cutoff contact map as an operator on the bosonic
Hilbert space.  In each fixed particle-number sector we prove that the
dressed contact map is bounded and that the renormalized principal
operator is invertible at sufficiently large negative energy.  The
resulting pseudo-resolvent reconstructs a unique self-adjoint sector
Hamiltonian, and the full number-conserving Hamiltonian is obtained as
their Hilbert-space direct sum.

The two-body sector provides an explicit illustration of the
construction.  Translation invariance permits decomposition at fixed
total momentum $P$, and the principal operator reduces to a scalar
function $\Phi_2(P,E)$ whose zeros determine the bound-pair dispersion.
The bound-state wave function follows naturally from the residue of the
renormalized resolvent.  At zero total momentum the resulting principal
function is directly related to that of the one-center
one-dimensional Salpeter delta problem; in the massless theory the
bound-pair dispersion can be obtained analytically.

An important consistency test is the nonrelativistic limit.  Restoring
$c$ and writing the physical two-body energy as
\begin{equation}
 E_B(c)=2mc^2-\varepsilon_B,
\end{equation}
with $\varepsilon_B>0$ fixed, the attractive Lieb--Liniger coupling is
\begin{equation}
 g_{\rm LL}
 =
 2\hbar\sqrt{\frac{\varepsilon_B}{m}}.
\end{equation}
Because the semirelativistic contact interaction must be renormalized
before the limit is taken, the natural order is
\[
 \text{regularization}
 \;\longrightarrow\;
 \text{renormalization}
 \;\longrightarrow\;
 \text{nonrelativistic limit}.
\]
We establish the corresponding sectorwise norm-resolvent limit directly
for the renormalized resolvents.

We also formulate a mean-field approximation after renormalization.
Rather than applying a Hartree ansatz to the cutoff-dependent Hamiltonian,
we use a product ansatz for the regular state in the one-orthofermion
sector.  In the massless theory, and more generally in the deeply bound
regime where the characteristic momenta satisfy
$p_{\rm char}\gg m$, the ultraviolet dispersion
$\omega(p)\simeq|p|$ controls the leading large-particle-number
behavior.  The resulting binding scale grows exponentially,
\begin{equation}
 |E_{\rm gr}^{\rm MF}|
 \sim
 \frac{\Lambda_B}{N}e^{cN},
 \qquad N=n-2\gg1,
\end{equation}
up to factors subleading relative to the exponential.  The coefficient
$c$ is determined by a one-dimensional variational problem related to a
fractional Gagliardo--Nirenberg inequality.

The remainder of the paper is organized as follows.
Section~\ref{sec:fock} introduces the second-quantized formulation and
the pair-annihilation operators.  Section~\ref{sec:heatkernel} develops
the Salpeter heat kernel and its short-time ultraviolet behavior.
Section~\ref{sec:principal} introduces the enlarged Fock space and the
Schur-complement construction.  Section~\ref{sec:renormalization}
performs the normal ordering and renormalization and treats the two-body
sector.  Section~\ref{sec:NR} identifies the nonrelativistic
Lieb--Liniger limit.  Section~\ref{sec:meanfield} develops the
renormalized mean-field approximation, and Sec.~\ref{sec:selfadjoint}
establishes existence and self-adjointness in every fixed particle-number
sector.  Appendices~\ref{app:normal} and \ref{app:estimates} contain the
normal-ordering and ultraviolet estimates, while
Appendix~\ref{app:NRnorm} gives the detailed sectorwise
norm-resolvent argument.

\section{Second-quantized formulation and notation}\label{sec:fock}

We use $\hbar=c=1$ until Sec.~\ref{sec:NR}.  Introduce bosonic creation and annihilation operators normalized by
\begin{equation}
 [a(p),a^\dagger(q)]=2\pi\delta(p-q),
 \qquad [a(p),a(q)]=0,
 \label{eq:CCR}
\end{equation}
and define
\begin{equation}
 \phi(x)=\int_{-\infty}^{\infty}\frac{\dd p}{2\pi}\,
 \ee^{\ii px}a(p),
 \qquad
 [\phi(x),\phi^\dagger(y)]=\delta(x-y).
 \label{eq:field}
\end{equation}

Strictly speaking, $a(p)$, $a^\dagger(p)$, $\phi(x)$, and
$\phi^\dagger(x)$ are operator-valued distributions and become
operators after smearing with suitable test functions.  We use the
standard unsmeared notation for brevity.  Hats distinguish
first-quantized observables from their numerical spectral variables:
$\hat p=-i\partial_x$, whereas $p$, $k$, and $P$ below are real
momentum variables.
The free Hamiltonian is the second quantization of the one-particle Salpeter operator $\hat h=\sqrt{\hat p^{\,2}+m^2}$,
\begin{equation}
 H_0=\int_{-\infty}^{\infty}\frac{\dd p}{2\pi}\,
 \w(p)a^\dagger(p)a(p),
 \qquad \w(p)=\sqrt{p^2+m^2}.
 \label{eq:H0}
\end{equation}
The formal interacting Hamiltonian is
\begin{equation}
 H=H_0-\frac{\lambda}{2}\int_{-\infty}^{\infty}\dd x\,
 \phi^\dagger(x)^2\phi(x)^2 .
 \label{eq:Hsecond}
\end{equation}
The factor $1/2$ is fixed by restriction to a sector of definite particle number.  Indeed, if
\begin{align}
 |\Psi_n\rangle
 &=\frac{1}{\sqrt{n!}}
 \int \dd x_1\cdots \dd x_n\,
 \Psi_n(x_1,\ldots,x_n)
 \nonumber\\
 &\qquad\times
 \phi^\dagger(x_1)\cdots\phi^\dagger(x_n)|0\rangle,
 \label{eq:nstate}
\end{align}
then Eq.~\eqref{eq:Hsecond} acts formally as Eq.~\eqref{eq:formalhamiltonian}.  The total boson number
\begin{equation}
 N_B=\int \dd x\,\phi^\dagger(x)\phi(x)
 \label{eq:number}
\end{equation}
is conserved.

It is useful to introduce a pair-annihilation operator $B(P)$, which removes two bosons whose momenta add to the fixed total momentum $P$,
\begin{equation}
 B(P)=\int_{-\infty}^{\infty}\frac{\dd k}{2\pi}\,
 a\left(\frac{P}{2}+k\right)
 a\left(\frac{P}{2}-k\right).
 \label{eq:Bpair}
\end{equation}
From Eq.~\eqref{eq:field},
\begin{equation}
 \phi(x)^2=\int\frac{\dd P}{2\pi}\,\ee^{\ii Px}B(P),
 \label{eq:pairFT}
\end{equation}
so that
\begin{equation}
 \int\dd x\,\phi^\dagger(x)^2\phi(x)^2
 =\int\frac{\dd P}{2\pi}\,B^\dagger(P)B(P).
 \label{eq:Vpair}
\end{equation}
Equation~\eqref{eq:Vpair} makes clear that the auxiliary degree of freedom introduced below should carry the total momentum of an interacting pair.

The pair momenta are $p_1=P/2+k$ and $p_2=P/2-k$, so
$p_1+p_2=P$ for every relative momentum $k$.  Equivalently, if
$\widehat P_{\rm tot}$ denotes the total momentum operator, then
\begin{equation}
 [\widehat P_{\rm tot},B(P)]=-P\,B(P).
\end{equation}
Thus the argument $P$ of $B(P)$ is the c-number total momentum removed
from the many-body state.

\section{Heat kernel of the one-particle free semirelativistic Hamiltonian}\label{sec:heatkernel}

The one-particle Salpeter heat kernel is
\begin{align}
 K_t(x-y)
 &=\mel{x}{\ee^{-t\sqrt{\hat p^{\,2}+m^2}}}{y}
 \nonumber\\
 &=\int_{-\infty}^{\infty}\frac{\dd p}{2\pi}\,
 \ee^{\ii p(x-y)}\ee^{-t\w(p)}.
 \label{eq:heatFT}
\end{align}
It is explicitly \cite{ErmanGadellaUncu2017,LiebLoss2001}
\begin{equation}
 K_t(r)=\frac{mt}{\pi\sqrt{t^2+r^2}}
 K_1\!\left(m\sqrt{t^2+r^2}\right),
 \label{eq:heatkernel}
\end{equation}
where $K_1$ is the modified Bessel function.  It obeys the semigroup identity
\begin{equation}
 \int_{-\infty}^{\infty}\dd z\,
 K_{t_1}(x-z)K_{t_2}(z-y)
 =K_{t_1+t_2}(x-y).
 \label{eq:semigroup}
\end{equation}
At coincidence, using $K_1(z)\sim z^{-1}$ as $z\to0^+$ \cite{Lebedev1972},
\begin{equation}
 K_t(0)=\frac{m}{\pi}K_1(mt)
 \sim \frac{1}{\pi t},
 \qquad t\to0^+.
 \label{eq:diagshort}
\end{equation}
For later estimates we record three useful norm identities or asymptotics,
\begin{align}
 \|K_t\|_{L^1(\R)}&=\ee^{-mt},
 \label{eq:L1K}\\
 \|K_t\|_{L^2(\R)}^2&=K_{2t}(0)
 \sim \frac{1}{2\pi t},
 \label{eq:L2K}\\
 \|K_t\|_{L^\infty(\R)}&=K_t(0)
 \sim \frac{1}{\pi t}.
 \label{eq:LinfK}
\end{align}
The first identity follows from the Fourier representation at zero momentum and positivity of the kernel; the second follows from Eq.~\eqref{eq:semigroup}.

\subsection{Distributional behavior of the squared kernel}

In the ultraviolet region $r\sim t\ll m^{-1}$,
\begin{equation}
 K_t(r)\sim \frac{1}{\pi}\frac{t}{t^2+r^2}.
 \label{eq:poisson}
\end{equation}
Let $f\in C_c^\infty(\R)$.  Then
\begin{align}
 \int\dd r\,K_t(r)^2f(r)
 &\sim \frac{1}{\pi^2}
 \int\dd r\,\frac{t^2f(r)}{(t^2+r^2)^2}
 \nonumber\\
 &=\frac{1}{\pi^2t}
 \int_{-\infty}^{\infty}\dd u\,
 \frac{f(tu)}{(1+u^2)^2}.
 \label{eq:distribution-step}
\end{align}
For each fixed $u$, $f(tu)\to f(0)$ as $t\to0^+$, while
\[
 \left|\frac{f(tu)}{(1+u^2)^2}\right|
 \leq \frac{\|f\|_\infty}{(1+u^2)^2},
\]
and the right-hand side is integrable.  Dominated convergence and
\begin{equation}
 \int_{-\infty}^{\infty}\frac{\dd u}{(1+u^2)^2}=\frac{\pi}{2}
 \label{eq:u-integral}
\end{equation}
therefore give
\begin{equation}
 K_t(r)^2\underset{t\to0^+}{\sim}
 \frac{1}{2\pi t}\delta(r)
 \label{eq:Ksquare}
\end{equation}
in the distributional sense.  Integrating Eq.~\eqref{eq:Ksquare} over $r$ gives the same coefficient as Eq.~\eqref{eq:L2K}.  Therefore the primitive two-body loop contains
\begin{equation}
 \int_0\dd t\,K_{2t}(0)
 \sim \frac{1}{2\pi}\int_0\frac{\dd t}{t},
 \label{eq:logdiv}
\end{equation}
which is logarithmically divergent.

\section{Extended Fock space and regularization}\label{sec:principal}

\subsection{Heat-kernel regularization}

Following the heat-kernel regularization used for many-body contact interactions \cite{ErmanTurgut2013}, define the smeared pair field
\begin{equation}
 \begin{split}
 B_\ep(y)=\int\dd x_1\dd x_2\,
 &K_\ep(x_1-y)K_\ep(y-x_2)
 \\[-1mm]
 &\times\phi(x_1)\phi(x_2).
 \end{split}
 \label{eq:Bepsx}
\end{equation}
Equivalently,
\begin{equation}
 B_\ep(P)=\int\frac{\dd k}{2\pi}\,
 \ee^{-\ep[\w(P/2+k)+\w(P/2-k)]}
 a(P/2+k)a(P/2-k).
 \label{eq:BepsP}
\end{equation}
As $\ep\to0^+$, $B_\ep(y)\to\phi(y)^2$ in the distributional sense.  The regularized Hamiltonian is
\begin{equation}
 H_\ep=H_0-\frac{\lambda(\ep)}{2}
 \int\dd y\,B_\ep^\dagger(y)B_\ep(y).
 \label{eq:Heps}
\end{equation}

\subsection{Continuous orthofermion}

Introduce auxiliary operators $\chi(x),\chi^\dagger(x)$ obeying the orthofermionic product algebra \cite{MishraRajasekaran1991}
\begin{equation}
 \chi(x)\chi^\dagger(y)=\delta(x-y)\Pi_0,
 \qquad
 \chi(x)\chi(y)=\chi^\dagger(x)\chi^\dagger(y)=0,
 \label{eq:ortho}
\end{equation}
where
\begin{equation}
 \Pi_1=\int\dd x\,\chi^\dagger(x)\chi(x),
 \qquad \Pi_0=1-\Pi_1.
 \label{eq:projectors}
\end{equation}
There is at most one orthofermion.  The enlarged space is
\begin{equation}
 \widetilde\F_B=\F_B\oplus[\F_B\otimes L^2(\R)].
 \label{eq:extendedF}
\end{equation}
Here
\begin{equation}
 \mathcal F_B=\bigoplus_{n=0}^{\infty}\mathcal H_n,
 \qquad
 \mathcal H_n=L^2_{\rm sym}(\mathbb R^n),
 \qquad \mathcal H_0\simeq\mathbb C.
\end{equation}
For fixed total physical particle number $n$, the relevant subspace is
\begin{equation}
 \widetilde\Hh_n=\Hh_n\oplus[\Hh_{n-2}\otimes L^2(\R)],
 \label{eq:fixedn}
\end{equation}
so one orthofermion represents two annihilated physical bosons.

Define
\begin{align}
 b_\ep&=\frac{1}{\sqrt2}\int\dd y\,
 \chi^\dagger(y)B_\ep(y),
 \label{eq:beps}\\
 b_\ep^\dagger&=\frac{1}{\sqrt2}\int\dd y\,
 B_\ep^\dagger(y)\chi(y).
 \label{eq:bepsdag}
\end{align}

Using $\chi(x)\chi^\dagger(y)=\delta(x-y)\Pi_0$,
\begin{equation}
 b_\epsilon^\dagger b_\epsilon
 =
 \frac12\int\dd y\,
 B_\epsilon^\dagger(y)B_\epsilon(y)\Pi_0.
\end{equation}
Hence, after restriction to the zero-orthofermion sector,
\begin{equation}
 H_\epsilon
 =
 H_0-\lambda(\epsilon)b_\epsilon^\dagger b_\epsilon.
\end{equation}
This identity is a compressed enlarged-space identity; it does not mean
that $b_\epsilon$ is an operator acting within the ordinary bosonic
space.
The purpose of the enlarged space is to linearize the quadratic pair
interaction and make the Schur-complement structure explicit.  The
orthofermion is not an additional physical particle: $b_\epsilon$ maps
an $n$-boson state to a state with $n-2$ ordinary bosons and one
auxiliary contact-pair coordinate.  This is summarized by the conserved
generalized particle-number operator
\begin{equation}
 \mathcal N=N_B+2\Pi_1,
 \qquad [\widetilde H_\epsilon,\mathcal N]=0.
\end{equation}

The augmented Hamiltonian is
\begin{equation}
 \widetilde H_\ep
 =H_0\Pi_0+b_\ep^\dagger+b_\ep
 +\frac{1}{\lambda(\ep)}\Pi_1.
 \label{eq:Htilde}
\end{equation}
With respect to the zero- and one-orthofermion sectors,
\begin{equation}
 \widetilde H_\ep-E\Pi_0
 =\begin{pmatrix}
 H_0-E & b_\ep^\dagger\\
 b_\ep & \lambda(\ep)^{-1}I
 \end{pmatrix}.
 \label{eq:block}
\end{equation}
For a block operator
\[
 M=\begin{pmatrix}A&B\\ C&D\end{pmatrix},
\]
the Schur complement with respect to an invertible $D$ is
$S_D=A-BD^{-1}C$, while the complementary Schur complement with respect
to an invertible $A$ is $S_A=D-CA^{-1}B$
\cite{HornJohnson2013}.  Applying the first of these formulas here gives
\begin{equation}
 H_0-E-b_\ep^\dagger\lambda(\ep)b_\ep
 =H_\ep-E,
 \label{eq:Schur1}
\end{equation}
where Eq.~\eqref{eq:ortho} was used.  The complementary Schur reduction, now with respect to the bosonic
block $H_0-E$, gives
\begin{equation}
 R_\ep(E)=R_0(E)+R_0(E)b_\ep^\dagger
 \Phi_\ep(E)^{-1}b_\ep R_0(E),
 \label{eq:resolventeps}
\end{equation}
with
\begin{equation}
 R_0(E)=(H_0-E)^{-1}
 \label{eq:R0}
\end{equation}
and regularized principal operator
\begin{equation}
 \Phi_\ep(E)=\frac{\Pi_1}{\lambda(\ep)}
 -b_\ep R_0(E)b_\ep^\dagger.
 \label{eq:Phieps}
\end{equation}
This is the direct analogue of the principal operator introduced for nonrelativistic many-body contact interactions \cite{Rajeev1999,ErmanTurgut2013}.

\section{Normal ordering and renormalization}\label{sec:renormalization}

The basic pull-through identity follows from
\begin{equation}
 H_0a^\dagger(p)=a^\dagger(p)[H_0+\w(p)],
 \label{eq:pull1}
\end{equation}
namely
\begin{equation}
 (H_0-E)^{-1}a^\dagger(p)
 =a^\dagger(p)[H_0-E+\w(p)]^{-1}.
 \label{eq:pull2}
\end{equation}
In coordinate space,
\begin{align}
 (H_0-E)^{-1}\phi^\dagger(x)
 &=\int\dd y\,\phi^\dagger(y)
 \int_0^\infty\dd t\,
 \nonumber\\[-1mm]
 &\qquad\times K_t(x-y)\ee^{-t(H_0-E)}.
 \label{eq:pullx}
\end{align}
For two creation operators,
\begin{align}
 &(H_0-E)^{-1}\phi^\dagger(x_1)\phi^\dagger(x_2)
 \nonumber\\
 &\quad=\int\dd y_1\dd y_2\,
 \phi^\dagger(y_1)\phi^\dagger(y_2)
 \int_0^\infty\dd t\,
 \nonumber\\[-1mm]
 &\qquad\times K_t(x_1-y_1)K_t(x_2-y_2)
 \ee^{-t(H_0-E)}.
 \label{eq:pulltwo}
\end{align}
Normal ordering the four field operators produces zero-, one-, and
two-contraction pieces.  Here a \emph{contraction} is simply the
commutator term generated when an annihilation field is moved past a
creation field,
\[
 \phi(x)\phi^\dagger(y)
 =
 \phi^\dagger(y)\phi(x)+\delta(x-y).
\]
Thus the terminology is the same one used in field-theoretic normal
ordering, but no vacuum expectation value is required in the present
algebraic step.  The explicit calculation is given in
Appendix~\ref{app:normal}.  The result is
\begin{equation}
 \Phi_\ep(E)=\frac{\Pi_1}{\lambda(\ep)}
 -\mathcal U_{2,\ep}(E)-\mathcal U_{1,\ep}(E)-\mathcal B_\ep(E),
 \label{eq:Phisplit-eps}
\end{equation}
where $\mathcal U_{2,\ep}$ is normal ordered and quartic in the bosonic fields, $\mathcal U_{1,\ep}$ is quadratic, and $\mathcal B_\ep$ contains no bosonic field operators.  The only ultraviolet divergence occurs in $\mathcal B_\ep$.

\subsection{Physical subtraction}

Let $E_B<2m$ denote the measured energy of the two-boson bound state at zero total momentum.  We choose the bare coupling by
\begin{equation}
 \frac{1}{\lambda(\ep)}
 =\int_0^\infty\dd t\,
 \ee^{tE_B}K_{2(t+2\ep)}(0).
 \label{eq:barecoupling-heat}
\end{equation}
Using Eq.~\eqref{eq:heatFT}, this is equivalently
\begin{equation}
 \frac{1}{\lambda(\ep)}
 =\int_{-\infty}^{\infty}\frac{\dd k}{2\pi}\,
 \frac{\ee^{-4\ep\w(k)}}{2\w(k)-E_B}.
 \label{eq:barecoupling-mom}
\end{equation}

To display the logarithm explicitly, set
$q=k/m$, $e_B=E_B/m$, and $a=4m\epsilon$.  Using
\[
 \frac{1}{2r-e_B}
 =
 \frac{1}{2r}+\frac{e_B}{2r(2r-e_B)},
 \qquad r=\sqrt{1+q^2},
\]
the second term is integrable at infinity, whereas the divergent part is
\begin{equation}
 \frac{1}{2\pi}\int_0^\infty\dd q\,
 \frac{e^{-a\sqrt{1+q^2}}}{\sqrt{1+q^2}}
 =
 \frac{1}{2\pi}K_0(4m\epsilon),
\end{equation}
after $q=\sinh u$.  Since
$K_0(z)=-\ln(z/2)-\gamma+O(z^2\ln z)$ as $z\to0^+$,
\begin{equation}
 \frac{1}{\lambda(\ep)}
 =\frac{1}{2\pi}\ln\frac{1}{m\epsilon}+O(1),
 \qquad \ep\to0^+,
 \label{eq:logrunning}
\end{equation}
where the finite $O(1)$ term depends on $E_B/m$ and on the precise regulator convention.  Equation~\eqref{eq:barecoupling-heat} is preferable because it implements the physical subtraction exactly rather than subtracting only the leading $1/t$ singularity.

\subsection{Renormalized principal operator}

After combining Eq.~\eqref{eq:barecoupling-heat} with the two-contraction term and taking $\ep\to0^+$, the renormalized principal operator can be written
\begin{equation}
 \Phi(E)=\Phi^{(0)}(E)+\Phi^{(1)}(E)+\Phi^{(2)}(E).
 \label{eq:Phiren}
\end{equation}
The zero-boson or pair-bubble part is
\begin{equation}
 \Phi^{(0)}(E)=
 \int\dd x\dd x'\,\chi^\dagger(x)
 \int_0^\infty\dd t\,
 \left[
 \ee^{tE_B}K_{2t}(0)\delta(x-x')
 -K_t(x-x')^2\ee^{-t(H_0-E)}
 \right]\chi(x').
 \label{eq:Phi0}
\end{equation}
The one-contraction term is
\begin{equation}
\begin{aligned}
 \Phi^{(1)}(E)={}&-2\int\dd x\dd x'\dd x_1\dd x_2\,
 \chi^\dagger(x)\phi^\dagger(x_1)
 \int_0^\infty\dd t\,
 K_t(x_1-x')K_t(x'-x)K_t(x-x_2)\\
 &\qquad\times\ee^{-t(H_0-E)}
 \phi(x_2)\chi(x')
 \end{aligned}
 \label{eq:Phi1}
\end{equation}
Finally,
\begin{equation}
\begin{aligned}
 \Phi^{(2)}(E)={}&-\frac12\int\dd x\dd x'\dd x_1\dd x_2\dd x_1'\dd x_2'\,
 \chi^\dagger(x)\phi^\dagger(x_1')\phi^\dagger(x_2')
 \int_0^\infty\dd t\,\\
 &\quad\times K_t(x_1'-x')K_t(x'-x_2')
 K_t(x_1-x)K_t(x-x_2)
 \ee^{-t(H_0-E)}
 \phi(x_1)\phi(x_2)\chi(x')
 \end{aligned}
 \label{eq:Phi2op}
\end{equation}
Equations~\eqref{eq:Phi0}--\eqref{eq:Phi2op} are the central
many-body result.  The bare zero-cutoff contact map is singular when
considered by itself.  The well-defined object is instead the
resolvent-dressed contact map
\begin{equation}
 T(E)
 :=
 \lim_{\epsilon\to0^+}b_\epsilon R_0(E),
 \qquad
 R_0(E)=(H_0-E)^{-1},
 \label{eq:contact-resolvent-map}
\end{equation}
where the limit is taken sectorwise.  The renormalized bosonic resolvent
is
\begin{equation}
 R(E)
 =
 R_0(E)
 +
 T(\overline E)^\dagger\Phi(E)^{-1}T(E).
 \label{eq:resolvent-ren}
\end{equation}
For real $E$ below the free threshold,
$T(\overline E)^\dagger=T(E)^\dagger$.  Thus no unsmeared contact
operator is required in the cutoff-free resolvent formula.

The candidate bound-state energies are determined by
\begin{equation}
 \Phi(E)|\Psi\rangle=0.
 \label{eq:boundprincipal}
\end{equation}
For fixed $n$, Eq.~\eqref{eq:boundprincipal} acts on $\Hh_{n-2}\otimes L^2(\R)$.

The subtraction in Eq.~\eqref{eq:Phi0} is locally finite because both terms have the same distributional short-time singularity,
\begin{equation}
 \ee^{tE_B}K_{2t}(0)\delta(x-x')
 \sim \frac{1}{2\pi t}\delta(x-x'),
 \label{eq:firstsing}
\end{equation}
while Eq.~\eqref{eq:Ksquare} gives the same leading term for the second contribution.  Appendix~\ref{app:estimates} explains in detail why the normal-ordered terms \eqref{eq:Phi1} and \eqref{eq:Phi2op} have a finite $\ep\to0^+$ limit and supplies the large-negative-energy bounds used in the existence proof.  Sec.~\ref{sec:selfadjoint} then adapts the pseudo-resolvent method of Ref.~\cite{DoganErmanTurgut2012} to show that the resulting resolvent defines a unique self-adjoint Hamiltonian in each fixed particle-number sector.

\subsection{Exact two-body sector}\label{sec:two-body}

For $n=2$ the one-orthofermion sector contains no ordinary bosons.  Consequently,
\begin{equation}
 \Phi^{(1)}(E)=\Phi^{(2)}(E)=0,
 \qquad H_0|0\rangle=0,
 \label{eq:twobodyvanish}
\end{equation}
and the principal operator is diagonal in the total momentum $P$ of the pair.  Defining
\begin{equation}
 \chi(P)=\int\dd x\,\ee^{-\ii Px}\chi(x),
 \qquad
 \chi(P)\chi^\dagger(Q)=2\pi\delta(P-Q)\Pi_0,
 \label{eq:chiP}
\end{equation}
we write
\begin{equation}
 \Phi(E)=\int\frac{\dd P}{2\pi}\,
 \chi^\dagger(P)\Phi_2(P,E)\chi(P).
 \label{eq:PhiPdiag}
\end{equation}
The Fourier transform of $K_t^2$ is
\begin{align}
 \int\dd r\,\ee^{-\ii Pr}K_t(r)^2
 &=\int\frac{\dd k}{2\pi}\,
 \ee^{-t\w(k)}\ee^{-t\w(P-k)}
 \nonumber\\
 &=\int\frac{\dd k}{2\pi}\,
 \ee^{-t[\w(P/2+k)+\w(P/2-k)]}.
 \label{eq:KsquareFT}
\end{align}
Also,
\begin{equation}
 K_{2t}(0)=\int\frac{\dd k}{2\pi}\,\ee^{-2t\w(k)}.
 \label{eq:K2tFT}
\end{equation}
Substitution into Eq.~\eqref{eq:Phi0} gives
\begin{align}
 \Phi_2(P,E)
 &=\int_0^\infty\dd t\int\frac{\dd k}{2\pi}
 \Big\{
 \ee^{-t[2\w(k)-E_B]}
 \nonumber\\
 &\qquad\qquad-
 \ee^{-t[\w(P/2+k)+\w(P/2-k)-E]}
 \Big\}.
 \label{eq:Phi2time}
\end{align}
For energies below the two-particle continuum threshold, the $t$ integration is elementary and yields
\begin{equation}
 \Phi_2(P,E)=\int_{-\infty}^{\infty}\frac{\dd k}{2\pi}
 \left[
 \frac{1}{2\w(k)-E_B}
 -\frac{1}{\w(P/2+k)+\w(P/2-k)-E}
 \right].
 \label{eq:Phi2exact}
\end{equation}
The two integrals in Eq.~\eqref{eq:Phi2exact} diverge logarithmically when considered separately, but their difference behaves as $O(k^{-2})$ and is convergent.

\subsection{Continuum threshold and monotonicity}

At fixed total momentum $P$, the free two-particle energy is
\begin{equation}
 \mathcal E_P(k)=\w(P/2+k)+\w(P/2-k).
 \label{eq:freepairenergy}
\end{equation}
Since $\w$ is convex, the minimum occurs at $k=0$, hence
\begin{equation}
 E_{\rm th}(P)=2\sqrt{m^2+P^2/4}=\sqrt{P^2+4m^2}.
 \label{eq:threshold}
\end{equation}
A bound state satisfies $E(P)<E_{\rm th}(P)$ and
\begin{equation}
\Phi_2(P,E(P))=0.
 \label{eq:bounddisp-general}
\end{equation}
Furthermore,
\begin{equation}
 \frac{\partial\Phi_2(P,E)}{\partial E}
 =-\int\frac{\dd k}{2\pi}\,
 \frac{1}{[\mathcal E_P(k)-E]^2}<0,
 \label{eq:monotonicity}
\end{equation}
so $\Phi_2(P,E)$ is strictly decreasing with $E$ below threshold.  Thus, whenever a zero exists at fixed $P$, it is unique.

\subsection{Bound-state wave function from the resolvent}

The singular contact interaction is defined here through the renormalized
resolvent rather than by treating the formal delta interaction as an
ordinary potential.  It is therefore natural to reconstruct a bound state
from the pole of the resolvent.  Let $E(P)$ be a simple zero of the
two-body principal function,
\begin{equation}
 \Phi_2(P,E(P))=0.
\end{equation}
Near the pole,
\begin{equation}
 \Phi_2(P,E)
 =
 (E-E(P))
 \left.
 \frac{\partial\Phi_2(P,E)}{\partial E}
 \right|_{E=E(P)}
 +O((E-E(P))^2).
\end{equation}
For the convention $R(E)=(H-E)^{-1}$, a normalized bound state contributes
\[
 R(E)\sim-\frac{|\Psi_P\rangle\langle\Psi_P|}{E-E(P)}.
\]
Comparing this residue with Eq.~\eqref{eq:resolvent-ren} gives, up to an
overall phase,
\begin{equation}
 |\Psi_P\rangle
 =
 \frac{
 T(E(P))^\dagger|P\rangle
 }{
 \sqrt{
 -\left.
 \dfrac{\partial\Phi_2(P,E)}{\partial E}
 \right|_{E=E(P)}
 }}.
 \label{eq:BoundStateFromResidue}
\end{equation}
Here $|P\rangle$ denotes the one-orthofermion contact-channel state of
total momentum $P$.

Writing
\begin{equation}
 |\Psi_P\rangle
 =
 \frac{1}{\sqrt2}\int\frac{\dd k}{2\pi}\,
 \widetilde\psi_P(k)
 a^\dagger(P/2+k)a^\dagger(P/2-k)|0\rangle,
\end{equation}
one obtains
\begin{equation}
 \widetilde\psi_P(k)=
 \frac{\mathcal N_P}{
 \omega(P/2+k)+\omega(P/2-k)-E(P)}.
 \label{eq:wfmomentum}
\end{equation}
Moreover,
\begin{equation}
 |\mathcal N_P|^{-2}
 =
 -\left.
 \frac{\partial\Phi_2(P,E)}{\partial E}
 \right|_{E=E(P)}.
 \label{eq:normprincipal}
\end{equation}
Thus both the form and normalization of the bound-state wave function
follow directly from the residue of the renormalized resolvent.

\subsection{Reduction to the single-center Salpeter problem}\label{sec:single-center}

At zero total momentum,
\begin{equation}
 \mathcal E_0(k)=2\w(k),
 \label{eq:P0pair}
\end{equation}
so Eq.~\eqref{eq:Phi2exact} becomes
\begin{align}
 \Phi_2(0,E)
 &=\int\frac{\dd k}{2\pi}
 \left[
 \frac{1}{2\w(k)-E_B}
 -\frac{1}{2\w(k)-E}
 \right]
 \nonumber\\
 &=\frac12\int\frac{\dd k}{2\pi}
 \left[
 \frac{1}{\w(k)-E_B/2}
 -\frac{1}{\w(k)-E/2}
 \right].
 \label{eq:P0relation}
\end{align}
The quantity in square brackets is exactly the renormalized one-center principal function of the one-dimensional Salpeter delta problem \cite{ErmanGadellaUncu2017}.  Thus
\begin{equation}
 \Phi_2^{\rm pair}(0,E;E_B)
 =\frac12\,
 \Phi_{\delta}^{(1)}\!\left(\frac{E}{2};\frac{E_B}{2}\right).
 \label{eq:exactmapping}
\end{equation}
This gives a nontrivial consistency check of the many-body construction.

For $|z|<m$, define
\begin{equation}
 F_m(z)=
 \frac{z}{\pi\sqrt{m^2-z^2}}
 \left[
 \frac{\pi}{2}
 +\tan^{-1}\!\left(\frac{z}{\sqrt{m^2-z^2}}\right)
 \right].
 \label{eq:Fm}
\end{equation}
The explicit one-center result then gives
\begin{equation}
 \Phi_2(0,E)=\frac12
 \left[F_m(E_B/2)-F_m(E/2)\right],
 \label{eq:Phi2P0closed}
\end{equation}
with analytic continuation used outside the displayed real domain.  The zero-momentum bound-state wave function is
\begin{equation}
 \widetilde\psi_0(k)=\frac{\mathcal N_0}{2\sqrt{k^2+m^2}-E_B}.
 \label{eq:wfp0}
\end{equation}

\subsection{Massless theory}\label{sec:massless}

Set $m=0$, so $\w(p)=|p|$.  Write the physical zero-momentum bound-state energy as
\begin{equation}
 E_B=-\kappa_B,
 \qquad \kappa_B>0.
 \label{eq:kappaB}
\end{equation}
Equation~\eqref{eq:Phi2exact} becomes
\begin{align}
 \Phi_2(P,E)
 &=\int\frac{\dd k}{2\pi}
 \left[
 \frac{1}{2|k|+\kappa_B}
 \right.
 \nonumber\\[-1mm]
 &\qquad\left.
 -\frac{1}{|k+P/2|+|k-P/2|-E}
 \right].
 \label{eq:masslessPhiStart}
\end{align}
Because the result is even in $P$, take $P\ge0$.  A bound state lies below the continuum edge $E=P$.  Define
\begin{equation}
 \Delta=P-E>0.
 \label{eq:Delta}
\end{equation}
Using
\begin{equation}
 |k+P/2|+|k-P/2|
 =\begin{cases}
 P,& |k|\le P/2,\\
 2|k|,& |k|\ge P/2,
 \end{cases}
 \label{eq:absidentity}
\end{equation}
and a symmetric cutoff $\Lambda>P/2$, the first term gives
\begin{equation}
 I_1(\Lambda)=\frac{1}{2\pi}
 \ln\frac{2\Lambda+\kappa_B}{\kappa_B},
 \label{eq:I1}
\end{equation}
whereas the second gives
\begin{equation}
 I_2(\Lambda)
 =\frac{P}{2\pi\Delta}
 +\frac{1}{2\pi}
 \ln\frac{2\Lambda-E}{\Delta}.
 \label{eq:I2}
\end{equation}
Taking the difference and sending $\Lambda\to\infty$ yields
\begin{equation}
 \Phi_2(P,E)=\frac{1}{2\pi}
 \left[
 \ln\frac{P-E}{\kappa_B}
 -\frac{P}{P-E}
 \right],\qquad P\ge0.
 \label{eq:masslessPhi}
\end{equation}
At $P=0$,
\begin{equation}
 \Phi_2(0,E)=\frac{1}{2\pi}\ln\frac{-E}{\kappa_B},
 \qquad E<0,
 \label{eq:masslessP0}
\end{equation}
so the subtraction indeed fixes the bound state at $E=-\kappa_B$.

\paragraph{Bound-pair dispersion.}

The bound-state equation $\Phi_2(P,E)=0$ gives
\begin{equation}
 \ln\frac{\Delta}{\kappa_B}=\frac{P}{\Delta}.
 \label{eq:masslessboundeq}
\end{equation}
Let $y=P/\Delta$.  Then
\begin{equation}
 y\ee^y=\frac{P}{\kappa_B},
 \label{eq:Lambertstep}
\end{equation}
so, using the Lambert $W$ function \cite{CorlessLambertW1996},
\begin{equation}
 y=W\!\left(\frac{P}{\kappa_B}\right).
 \label{eq:LambertW}
\end{equation}
Therefore
\begin{equation}
 E(P)=P-\frac{P}{W(P/\kappa_B)},\qquad P\ge0.
 \label{eq:masslessdisp}
\end{equation}
Restoring parity, replace $P$ by $|P|$ on the right-hand side.  Using
\begin{equation}
 W(z)=z-z^2+\frac32z^3+O(z^4),
 \label{eq:Wseries}
\end{equation}
we find
\begin{equation}
 E(P)=-\kappa_B+\frac{P^2}{2\kappa_B}
 +O\!\left(\frac{|P|^3}{\kappa_B^2}\right).
 \label{eq:lowPmassless}
\end{equation}
Thus two massless constituents form a bound composite with a quadratic low-momentum dispersion controlled by the dynamically generated binding scale.

\section{Nonrelativistic limit}
\label{sec:NR}

We now restore $c$ and $\hbar$ and identify the ordinary
Lieb--Liniger theory obtained at low velocities.  The one-particle
dispersion is
\begin{equation}
 \omega_c(p)
 =
 \sqrt{m^2c^4+c^2p^2}
 =
 mc^2+\frac{p^2}{2m}
 -\frac{p^4}{8m^3c^2}
 +O(c^{-4}).
 \label{eq:NRexpansion}
\end{equation}
Thus, after subtraction of the rest energy $mc^2$ per particle,
\begin{equation}
 \sqrt{m^2c^4+c^2\hat p^{\,2}}-mc^2
 \longrightarrow
 \frac{\hat p^{\,2}}{2m},
 \qquad c\to\infty.
\end{equation}

At the formal level the limiting $n$-boson Hamiltonian is
\begin{equation}
 H_n^{\rm LL}
 =
 \sum_{\alpha=1}^{n}
 \frac{\hat p_\alpha^{\,2}}{2m}
 -
 g_{\rm LL}
 \sum_{\alpha<\beta}
 \delta(x_\alpha-x_\beta),
 \qquad g_{\rm LL}>0.
 \label{eq:LL}
\end{equation}
To obtain a finite nonrelativistic two-body binding energy, write
\begin{equation}
 E_B(c)=2mc^2-\varepsilon_B,
 \qquad \varepsilon_B>0,
 \label{eq:EBNR}
\end{equation}
and keep $\varepsilon_B$ fixed.  Then
\begin{equation}
 2\omega_c(k)-E_B(c)
 =
 \frac{k^2}{m}+\varepsilon_B+O(c^{-2}).
 \label{eq:denNR}
\end{equation}
If the energy at which the two-body principal function is evaluated is
written as $E=2mc^2-\varepsilon$, its zero-total-momentum limit is
\begin{equation}
 \Phi_2(0,E)
 \longrightarrow
 \int_{-\infty}^{\infty}\frac{\dd k}{2\pi\hbar}
 \left[
 \frac{1}{k^2/m+\varepsilon_B}
 -
 \frac{1}{k^2/m+\varepsilon}
 \right].
 \label{eq:PhiNR}
\end{equation}

For two equal-mass particles, the relative attractive-delta Hamiltonian
is
\begin{equation}
 H_{\rm rel}
 =
 \frac{\hat p^{\,2}}{m}
 -
 g_{\rm LL}\delta(x),
\end{equation}
whose unique bound state has
\begin{equation}
 \varepsilon_B
 =
 \frac{m g_{\rm LL}^2}{4\hbar^2}.
\end{equation}
Consequently,
\begin{equation}
 g_{\rm LL}
 =
 2\hbar\sqrt{\frac{\varepsilon_B}{m}},
 \qquad
 \varepsilon_B=2mc^2-E_B(c).
 \label{eq:matchingLL}
\end{equation}
The zero-total-momentum wave function correspondingly reduces to
\begin{equation}
 \widetilde\psi_0(k)
 \propto
 \frac{1}{k^2/m+\varepsilon_B},
\end{equation}
as expected for the one-dimensional attractive delta interaction.

The preceding argument identifies the limiting dispersion, subtraction
condition, coupling, and two-body bound state.  The stronger operator
statement requires the limit to be taken only after the
semirelativistic interaction has been renormalized:
\[
 \text{regularization}
 \;\longrightarrow\;
 \text{renormalization}
 \;\longrightarrow\;
 \text{nonrelativistic limit}.
\]
Appendix~\ref{app:NRnorm} implements this order directly at the
resolvent level and proves, for every fixed $n$,
\begin{equation}
 H_{n,c}-nmc^2
 \xrightarrow[c\to\infty]{\mathrm{norm\ resolvent}}
 H_n^{\rm LL}.
 \label{eq:NR-main-norm}
\end{equation}

\section{Mean-field approximation}
\label{sec:meanfield}

The renormalized principal-operator formulation suggests a mean-field approximation that is more natural for the present singular problem than applying a Hartree ansatz directly to the bare Hamiltonian.  The reason is that the physical $n$-boson bound-state wave function reconstructed from the resolvent contains the short-distance contact-pair convolution and therefore need not approach a simple product state even when $n$ is large.  By contrast, the eigenvector of the well-defined principal operator lives in the regular one-orthofermion sector and can be approximated by a product state.  This is the strategy used for the nonrelativistic many-body point interaction in Ref.~\cite{ErmanTurgut2013}.  We first formulate the corresponding principal-operator mean field and then obtain a large-$n$ estimate of the binding energy.  A brief effective-coupling comparison is included at the end.

\subsection{Product ansatz in the one-orthofermion sector}

For a fixed physical particle number $n$, the principal operator acts on
\begin{equation}
 \mathcal H_{n-2}\otimes L^2(\R).
 \label{eq:MFPOspace}
\end{equation}
Throughout this section we write
\begin{equation}
 N=n-2,
 \label{eq:MFN}
\end{equation}
so that $n$ denotes the total physical particle number and $N$ the number of ordinary bosons remaining in the one-orthofermion sector.  Let
\begin{equation}
 a_u^\dagger=\int_{\R}\dd x\,u(x)\phi^\dagger(x),
 \qquad
 \int_{-\infty}^{\infty} \dd x\,|u(x)|^2=1,
 \label{eq:MFPOau}
\end{equation}
and introduce a normalized orthofermion wave function
\begin{equation}
 \int_{-\infty}^{\infty} \dd x\,|\psi(x)|^2=1.
 \label{eq:MFPOnormpsi}
\end{equation}
The mean-field trial state for the lowest eigenvector of the principal operator is taken to be
\begin{equation}
 |\Theta^{(n)}_{u,\psi}\rangle
 =\frac{(a_u^\dagger)^N}{\sqrt{N!}}|0\rangle_B
 \otimes
 \int_{-\infty}^{\infty} \dd x\,\psi(x)\chi^\dagger(x)|0\rangle_\chi .
 \label{eq:MFPOansatz}
\end{equation}
The function $u(x)$ describes the common orbital of the $N$ remaining
ordinary bosons in the one-orthofermion sector, whereas $\psi(x)$ describes
the position of the auxiliary contact pair.  The latter is not an additional
physical particle; rather, it represents the two bosons that have been
removed at the contact vertex and replaced by a single auxiliary degree of
freedom.

For a self-bound state on the full line, a localized product ansatz selects a center of the cluster and thus breaks translation invariance at the mean-field level.  This is the usual situation for a self-bound Hartree state; translational invariance of the exact state can, if desired, be restored by superposing translated copies of the localized solution.

\subsection{Heat-evolved orbital and useful quantities}

Let
\begin{equation}
 \hat h=\sqrt{\hat p^{\,2}+m^2}
 \label{eq:MFPOh}
\end{equation}
be the one-particle Salpeter Hamiltonian and define
\begin{equation}
 u_t(x)=\langle x|e^{-t\hat h}|u\rangle=\left(e^{-t\hat h}u\right)(x)
 =\int_{-\infty}^{\infty} \dd y\,K_t(x-y)u(y).
 \label{eq:MFPOut}
\end{equation}
We also introduce
\begin{equation}
 A_u(t)=\langle u|e^{-t\hat h}|u\rangle
 =\int_{-\infty}^{\infty} \dd x\,u^*(x)u_t(x).
 \label{eq:MFPOA}
\end{equation}

On the $N$-boson factor, the free Hamiltonian \eqref{eq:H0} reduces to
\begin{equation}
 H_0^{(N)}
 =
 \left.d\Gamma(\hat h)\right|_{\mathcal H_N}
 =
 \sum_{j=1}^{N}\hat h_j .
\end{equation}
Since $[\hat h_i,\hat h_j]=0$ for $i\neq j$,
\begin{equation}
 e^{-tH_0^{(N)}}=(e^{-t\hat h})^{\otimes N}.
\end{equation}
Hence, for the Hartree product $u^{\otimes N}$,
\begin{align}
\langle u^{\otimes N}|e^{-tH_0^{(N)}}|u^{\otimes N}\rangle
&=
\prod_{j=1}^{N}\langle u|e^{-t\hat h}|u\rangle
\nonumber\\
&=[A_u(t)]^N.  \label{eq:MFPOsemigroup}
\end{align}
Thus the free propagation of the $N$ bosons is reduced to the $N$th power
of a single one-particle matrix element.

For later use we define
\begin{align}
 B_\psi(t)
 &=\int_{\R^2}\dd x\dd x'\,
 \psi^*(x)K_t(x-x')^2\psi(x'),
 \label{eq:MFPOB}\\
 C_{u,\psi}(t)
 &=\int_{\R^2}\dd x\dd x'\,
 \psi^*(x)\psi(x')u_t(x)
 K_t(x-x')u_t^*(x'),
 \label{eq:MFPOC}\\
 D_{u,\psi}(t)
 &=\int_{-\infty}^{\infty} \dd x\,\psi^*(x)[u_t(x)]^2.
 \label{eq:MFPOD}
\end{align}
The quantity $B_\psi$ describes the free propagation of the contact-pair
degree of freedom, $C_{u,\psi}$ describes its coupling to one of the
remaining ordinary bosons, and $D_{u,\psi}$ describes the contribution
involving two of the remaining ordinary bosons.

\subsection{Expectation value of the renormalized principal operator}

The renormalized operator derived in Sec.~\ref{sec:renormalization} has the decomposition
\begin{equation}
 \Phi(E)=\Phi^{(0)}(E)+\Phi^{(1)}(E)+\Phi^{(2)}(E).
 \label{eq:MFPOPhi}
\end{equation}
Its expectation value in the state \eqref{eq:MFPOansatz} can be evaluated exactly within the product ansatz.

For the pair-bubble part, Eq.~\eqref{eq:Phi0} gives
\begin{align}
 \mathcal F_n^{(0)}(E;u,\psi)
 &\equiv
 \langle\Theta^{(n)}_{u,\psi}|\Phi^{(0)}(E)|\Theta^{(n)}_{u,\psi}\rangle
 \nonumber\\
 &=\int_0^\infty\dd t\,
 \left[
 e^{tE_B}K_{2t}(0)
 -e^{tE}[A_u(t)]^N B_\psi(t)
 \right].
 \label{eq:MFPOF0}
\end{align}
The important point is that this expression already contains the physical two-body scale $E_B$; no bare coupling appears.

For the one-contraction term we use
\begin{equation}
 \phi(x)|u^{\otimes N}\rangle
 =\sqrt{N}\,u(x)|u^{\otimes(N-1)}\rangle,
 \label{eq:MFPOann1}
\end{equation}
which yields
\begin{align}
 &\langle u^{\otimes N}|\phi^\dagger(x_1)e^{-tH_0}\phi(x_2)|u^{\otimes N}\rangle
 =N\,u^*(x_1)u(x_2)[A_u(t)]^{N-1}.
 \label{eq:MFPOonebody}
\end{align}
Substitution into Eq.~\eqref{eq:Phi1} gives
\begin{equation}
 \mathcal F_n^{(1)}(E;u,\psi)
 =-2N\int_0^\infty\dd t\,
 e^{tE}[A_u(t)]^{N-1}C_{u,\psi}(t).
 \label{eq:MFPOF1}
\end{equation}
Similarly,
\begin{align}
 &\langle u^{\otimes N}|\phi^\dagger(x_1')\phi^\dagger(x_2')e^{-tH_0}
 \phi(x_1)\phi(x_2)|u^{\otimes N}\rangle
 \nonumber\\
 &\qquad=N(N-1)u^*(x_1')u^*(x_2')u(x_1)u(x_2)
 [A_u(t)]^{N-2},
 \label{eq:MFPOtwobody}
\end{align}
so Eq.~\eqref{eq:Phi2op} becomes
\begin{equation}
 \mathcal F_n^{(2)}(E;u,\psi)
 =-\frac{N(N-1)}{2}\int_0^\infty\dd t\,
 e^{tE}[A_u(t)]^{N-2}|D_{u,\psi}(t)|^2.
 \label{eq:MFPOF2}
\end{equation}
In the matrix elements above, $H_0$ is understood on the
finite-particle sector that remains after the annihilation fields have
acted: the $(N-1)$-boson sector in the one-contraction term and the
$(N-2)$-boson sector in the zero-contraction term.
Combining the three pieces, the principal-operator mean-field functional is
\begin{align}
 \mathcal F_n(E;u,\psi)
 &\equiv
 \langle\Theta^{(n)}_{u,\psi}|\Phi(E)|\Theta^{(n)}_{u,\psi}\rangle
 \nonumber\\
 &=\int_0^\infty\dd t\,\Bigg\{
 e^{tE_B}K_{2t}(0)
 -e^{tE}[A_u(t)]^N B_\psi(t)
 \nonumber\\
 &\qquad
 -2N e^{tE}[A_u(t)]^{N-1}C_{u,\psi}(t)
 \nonumber\\
 &\qquad
 -\frac{N(N-1)}{2}e^{tE}[A_u(t)]^{N-2}|D_{u,\psi}(t)|^2
 \Bigg\}.
 \label{eq:MFPOfunctional}
\end{align}
This is the basic mean-field equation of the renormalized theory.

The relation between the principal operator and the physical spectrum can be
understood by following its eigenvalues as the energy is varied. 
Let $\omega_{0,n}(E)$ denote the lowest eigenvalue of the principal operator
$\Phi_n(E)$.  By the Rayleigh--Ritz variational principle,
\begin{equation}
\omega_{0,n}(E)
=
\inf_{\substack{|\Psi\rangle\in
\mathcal H_{n-2}\otimes L^2(\mathbb R)\\
\|\Psi\|=1}}
\langle\Psi|\Phi_n(E)|\Psi\rangle .
\label{eq:lowestPhi}
\end{equation}
At finite
cutoff one has
\begin{equation}
 \frac{\partial\Phi_{n,\epsilon}(E)}{\partial E}
 =
 -b_\epsilon R_0(E)^2 b_\epsilon^\dagger .
 \label{eq:PhiMonotoneMF}
\end{equation}
Consequently, for any state $|\Psi\rangle$ in the one-orthofermion sector,
\begin{equation}
 \left\langle\Psi\left|
 \frac{\partial\Phi_{n,\epsilon}(E)}{\partial E}
 \right|\Psi\right\rangle
 =
 -\left\|R_0(E)b_\epsilon^\dagger|\Psi\rangle\right\|^2
 \leq0 .
 \label{eq:PhiMonotoneFormMF}
\end{equation}
Thus the principal operator decreases as $E$ increases.  The same
monotonicity is retained after the cutoff is removed and the renormalized
principal operator is obtained.

Hence, by the Feynman--Hellmann theorem,
\begin{equation}
 \frac{d\omega_{k,n}(E)}{dE}
 =
 \left\langle\varphi_{k,n}(E),
 \frac{\partial\Phi_n(E)}{\partial E}
 \varphi_{k,n}(E)\right\rangle
 \leq0.
 \label{eq:eigenflowMF}
\end{equation}
Thus the eigenvalues of the principal operator decrease as $E$ increases.
Under the usual assumption that the bottom of the spectrum is a discrete
eigenvalue, the exact $n$-body ground-state energy is determined by
\begin{equation}
 \omega_{0,n}(E_{\rm gr}^{(n)})=0.
 \label{eq:exactgroundPhiMF}
\end{equation}

The mean-field ansatz restricts the variational space to the product states
$|\Theta_{u,\psi}^{(n)}\rangle$.  We therefore define the restricted
Rayleigh--Ritz value
\begin{equation}
 \Omega_n(E)
 =
 \inf_{\substack{\|u\|_2=1\\ \|\psi\|_2=1}}
 \mathcal F_n(E;u,\psi),
 \qquad
 \mathcal F_n(E;u,\psi)
 =
 \langle\Theta_{u,\psi}^{(n)}
 |\Phi_n(E)|
 \Theta_{u,\psi}^{(n)}\rangle .
 \label{eq:MFPOomega}
\end{equation}
Since the minimization is performed over a restricted class of states,
\begin{equation}
 \Omega_n(E)\geq\omega_{0,n}(E).
 \label{eq:MFvariationalbound}
\end{equation}
The principal-operator mean-field estimate $E_{\rm gr}^{\rm MF}$ is then
defined, provided such a zero exists, by
\begin{equation}
 \Omega_n(E_{\rm gr}^{\rm MF})=0.
 \label{eq:MFPOground}
\end{equation}
Because $\Omega_n(E)$ inherits the monotonic decrease with $E$, this gives
the variational bound
\begin{equation}
 E_{\rm gr}^{(n)}\leq E_{\rm gr}^{\rm MF}.
 \label{eq:MFenergybound}
\end{equation}
Equality is obtained only if the exact lowest eigenvector of the principal
operator belongs to the chosen mean-field variational family.

For fixed $u$ and $E$, the functional is quadratic in $\psi$ and can be written
\begin{equation}
 \mathcal F_n(E;u,\psi)=\langle\psi|\mathcal M_{u,E}|\psi\rangle,
 \label{eq:MFPOkerneldef}
\end{equation}
with kernel
\begin{align}
 \mathcal M_{u,E}(x,x')
 =\int_0^\infty\dd t\,\Bigg\{&
 e^{tE_B}K_{2t}(0)\delta(x-x')
 -e^{tE}A_u(t)^N K_t(x-x')^2
 \nonumber\\
 &-2N e^{tE}A_u(t)^{N-1}
 u_t(x)K_t(x-x')u_t^*(x')
 \nonumber\\
 &-\frac{N(N-1)}{2}e^{tE}A_u(t)^{N-2}
 u_t(x)^2u_t^*(x')^2
 \Bigg\}.
 \label{eq:MFPOkernel}
\end{align}
Thus, for a given condensate orbital, the contact-pair wave function satisfies an effective integral equation.  At the mean-field bound-state energy its lowest eigenvalue is zero,
\begin{equation}
 \int_{-\infty}^{\infty}\dd x'\,\mathcal M_{u,E_{\rm gr}^{\rm MF}}(x,x')\psi(x')=0.
 \label{eq:MFPOpaireq}
\end{equation}
This gives a useful physical picture: the auxiliary contact pair propagates in a medium formed by the remaining $N=n-2$ ordinary bosons, while the orbital $u$ must itself be determined self-consistently in the presence of that pair.

\subsection{Large-$n$ estimate of the binding energy}
\label{subsec:MFPOlargeN}

\paragraph{Massless case.}

The mean-field functional also permits an analytic estimate of the deeply bound large-$n$ spectrum.  The cleanest derivation is obtained for the massless model, for which
\begin{equation}
 \w(p)=|p|,
 \qquad
 E_B=-\kappa_B,
 \qquad
 E=-\kappa,
 \qquad
 \kappa,\kappa_B>0.
 \label{eq:MFPOmasslessdefs}
\end{equation}
The same leading ultraviolet analysis applies to the massive theory when the binding scale is much larger than $m$.

We first define the one-particle massless kinetic-energy functional
\begin{equation}
 \mathcal K[u]
 =
 \langle u||\hat p||u\rangle
 =
\int_{-\infty}^{\infty}\frac{\dd p}{2\pi}\,
 |p|\,|\widetilde u(p)|^2
 \label{eq:MFPOK}
\end{equation}
and
\begin{equation}
 \mathcal Q[u]
 =
 \int_{-\infty}^{\infty} \dd x\,|u(x)|^4.
 \label{eq:MFPOQ}
\end{equation}
Since, in the massless theory, the one-particle Hamiltonian is
$h=|\hat p|$, we have
\begin{equation}
 A_u(t)
 =
 \langle u|e^{-t|\hat p|}|u\rangle
 =
 \int_{-\infty}^{\infty} \frac{\dd p}{2\pi}\,
 e^{-t|p|}|\widetilde u(p)|^2 .
 \label{eq:AuMomentum}
\end{equation}
For a normalized orbital,
$\int_{-\infty}^{\infty} \frac{\dd p}{2\pi}|\widetilde u(p)|^2=1$. Expanding the exponential for $t\to0^+$ therefore gives
\begin{align}
 A_u(t)
 &=
 \int_{-\infty}^{\infty} \frac{\dd p}{2\pi}
 \left[1-t|p|+O(t^2p^2)\right]
 |\widetilde u(p)|^2
 \nonumber\\
 &=
 1-t\mathcal K[u]+O(t^2),
 \label{eq:AuShortDerivation}
\end{align}
where the displayed $O(t^2)$ form assumes a finite second momentum moment.
Consequently,
\begin{align}
 [A_u(t)]^N
 &=
 \exp\!\left[N\ln A_u(t)\right]
 \nonumber\\
 &=
 \exp\!\left[-Nt\mathcal K[u]+O(Nt^2)\right]
 \simeq
 e^{-Nt\mathcal K[u]} .
 \label{eq:AuPowerShort}
\end{align}
The last approximation is valid in the short-time region relevant to the
deeply bound large-$N$ regime.  Indeed, the subsequent heat-kernel integrals
are dominated by
$t\sim[\kappa+N\mathcal K[u]]^{-1}$, which becomes small as the binding
scale increases.

Moreover, the distributional short-time relation $K_t(x)^2\sim(2\pi t)^{-1}\delta(x)$ gives
\begin{equation}
 B_\psi(t)\simeq\frac{1}{2\pi t}.
 \label{eq:MFPOBshort}
\end{equation}
The pair-bubble part therefore becomes
\begin{align}
 \mathcal F_n^{(0)}
 &\simeq
 \frac{1}{2\pi}\int_0^\infty\frac{\dd t}{t}
 \left[e^{-\kappa_B t}
 -e^{-[\kappa+N\mathcal K[u]]t}\right]
 \nonumber\\
 &=\frac{1}{2\pi}
 \ln\!\left(\frac{\kappa+N\mathcal K[u]}{\kappa_B}\right).
 \label{eq:MFPOF0asymp}
\end{align}

The term with the largest combinatorial factor is
$\mathcal F_n^{(2)}$.  At short times,
\begin{equation}
 D_{u,\psi}(t)
 =
 D_{u,\psi}(0)+O(t),
 \qquad
 D_{u,\psi}(0)
 =
 \int_{-\infty}^{\infty} \dd x\,\psi^*(x)u(x)^2 .
 \label{eq:MFPOD0}
\end{equation}
For normalized $\psi$, the Cauchy--Schwarz inequality gives
\begin{equation}
 |D_{u,\psi}(0)|^2
 \leq
 \|\psi\|_2^2\,\|u^2\|_2^2
 =
 \mathcal Q[u].
 \label{eq:MFPODCS}
\end{equation}
The upper bound is saturated when the pair wave function is chosen as
\begin{equation}
 \psi(x)
 =
 \frac{u(x)^2}{\sqrt{\mathcal Q[u]}} ,
 \label{eq:MFPOpsiopt}
\end{equation}
up to an irrelevant overall phase.  Thus
$|D_{u,\psi}(0)|^2=\mathcal Q[u]$.

For completeness, the first short-time correction follows from
$u_t=e^{-t|\hat p|}u=u-t|\hat p|u+O(t^2)$ and is
\begin{equation}
 |D_{u,\psi}(t)|^2
 =
 \mathcal Q[u]
 -
 4t\,\mathcal I[u]
 +
 O(t^2),
 \qquad
 \mathcal I[u]
 =
 \int_{-\infty}^{\infty} \dd x\,u(x)^3|\hat p|u(x),
 \label{eq:MFPODnext}
\end{equation}
where $u$ has been taken real.  

The choice~\eqref{eq:MFPOpsiopt} should be understood as the leading
large-$N$ variational choice for the pair wave function.  Indeed,
$\mathcal F_n^{(2)}$ is the dominant $\psi$-dependent contribution, whereas
the one-contraction term is subleading.  To see this explicitly, the
short-time limit of $C_{u,\psi}(t)$ is
\begin{equation}
 C_{u,\psi}(t)
 =
 C_{u,\psi}(0)+O(t),
 \qquad
 C_{u,\psi}(0)
 =
 \int_{-\infty}^{\infty} \dd x\,|\psi(x)|^2|u(x)|^2 .
 \label{eq:MFPOCshort}
\end{equation}
For the choice
$\psi=u^2/\sqrt{\mathcal Q[u]}$ this becomes
\begin{equation}
 C_{u,\psi}(0)
 =
 \mathcal C[u]
 \equiv
 \frac{\displaystyle \int_{-\infty}^{\infty} \dd x\,|u(x)|^6}
 {\displaystyle\mathcal Q[u]} .
 \label{eq:MFPOCzero}
\end{equation}
Using $[A_u(t)]^{N-1}\simeq
e^{-(N-1)t\mathcal K[u]}$, we therefore find
\begin{equation}
 \mathcal F_n^{(1)}
 \simeq
 -\frac{2N\mathcal C[u]}
 {\kappa+(N-1)\mathcal K[u]}.
 \label{eq:MFPOF1asymp}
\end{equation}
For the scaled orbital $u_a(x)=a^{1/2}v(ax)$,
$\mathcal C[u_a]\propto a$, while
$\mathcal K[u_a]\propto a$.  At the optimized large-$N$ scale,
$\kappa+N\mathcal K[u_a]\sim Na$, and hence
\begin{equation}
 \mathcal F_n^{(1)}=O(1),
 \qquad
 \mathcal F_n^{(2)}=O(N).
 \label{eq:MFPOorders}
\end{equation}
Thus $\mathcal F_n^{(1)}$ is suppressed by one power of $N$ and does not
modify the leading exponential large-$N$ binding law.  A systematic
next-to-leading-order treatment would require keeping this term together
with the $O(t)$ correction in $D_{u,\psi}(t)$ and the corresponding
higher-order terms in $A_u(t)$.

Thus, to leading order in $N$,
\begin{equation}
 \mathcal F_n^{(2)}
 \simeq
 -\frac{N(N-1)\mathcal Q[u]}
 {2[\kappa+(N-2)\mathcal K[u]]}
 =
 -\frac{N^2\mathcal Q[u]}
 {2[\kappa+N\mathcal K[u]]}
 \left[1+O\!\left(\frac1N\right)\right].
 \label{eq:MFPOF2asymp}
\end{equation}

Since corrections from the next term in
$[A_u(t)]^{N-2}$ and from $\mathcal F_n^{(1)}$ enter at the same order, we
consistently retain only the leading short-time contribution in the
large-$N$ estimate below.

The zero-eigenvalue condition then reduces to
\begin{equation}
 \ln\!\left(\frac{\kappa+N\mathcal K[u]}{\kappa_B}\right)
 \simeq
 \frac{\pi N^2\mathcal Q[u]}
 {\kappa+N\mathcal K[u]}.
 \label{eq:MFPOlargeNeq}
\end{equation}
The logarithm on the left is the direct manifestation of the marginal ultraviolet interaction.  It is this logarithm that leads to an exponentially large many-body binding scale.

To make the scaling explicit, write a normalized orbital as
\begin{equation}
 u_a(x)=a^{1/2}v(ax),
 \qquad \|v\|_2=1.
 \label{eq:MFPOscaling}
\end{equation}
For the massless kinetic operator,
\begin{equation}
 \mathcal K[u_a]=a\mathcal K[v],
 \qquad
 \mathcal Q[u_a]=a\mathcal Q[v].
 \label{eq:MFPOscaleKQ}
\end{equation}

Thus the dimensionless ratio
\begin{equation}
 r[v]
 =
 \frac{\mathcal Q[v]}{\mathcal K[v]}
 =
 \frac{\mathcal Q[u_a]}{\mathcal K[u_a]}
 \label{eq:MFPOr}
\end{equation}
depends only on the shape $v$ and not on the scale $a$ of the trial
orbital.  To perform the remaining scale optimization, we introduce
\begin{equation}
 z=\frac{N\mathcal K[u_a]}{\kappa}.
 \label{eq:MFPOz}
\end{equation}
This variable measures the ratio of the total one-particle kinetic scale to
the binding scale.  Since
$N\mathcal K[u_a]=z\kappa$, and $\mathcal Q[u_a]
=
r[v]\mathcal K[u_a]
=
r[v]\frac{z\kappa}{N}$, the leading zero-eigenvalue condition becomes
\begin{equation}
 \ln\left[\frac{\kappa(1+z)}{\kappa_B}\right]
 \simeq
 \pi r[v]N\frac{z}{1+z}.
\end{equation}
Solving for $\kappa$ gives
\begin{equation}
 \kappa(z)
 \simeq
 \frac{\kappa_B}{1+z}
 \exp\left[
 \pi r[v]N\frac{z}{1+z}
 \right].
 \label{eq:MFPOkappaz}
\end{equation}

For a fixed shape $v$, changing the scale $a$ is equivalent to changing
$z$.  Since $E=-\kappa$, the best binding estimate is obtained by maximizing
$\kappa(z)$.  Differentiating its logarithm gives
\[
 \frac{d}{dz}\ln\kappa(z)
 =
 -\frac{1}{1+z}
 +
 \frac{\pi r[v]N}{(1+z)^2},
\]
and hence the stationary point satisfies
\begin{equation}
 1+z_*=\pi r[v]N.
 \label{eq:MFPOzstar}
\end{equation}
For $N\gg1$ this point gives the maximum.  Substitution back into
Eq.~\eqref{eq:MFPOkappaz} yields
\begin{equation}
 \kappa_{\rm MF}[v]
 \simeq
 \frac{\kappa_B}
 {e\,\pi r[v]N}
 \exp\!\bigl(\pi r[v]N\bigr).
 \label{eq:MFPOkappatrial}
\end{equation}
Thus, defining $c[v]=\pi r[v]$, the leading large-$N$ behavior is
\begin{equation}
 |E_{\rm gr}^{\rm MF}|
 \sim
 \frac{\kappa_B}{N}e^{cN},
 \qquad
 c=\pi r,
 \qquad N=n-2\gg1,
 \label{eq:MFPOexpbinding}
\end{equation}
where factors subleading relative to the exponential have been suppressed.

The same qualitative exponential behavior was found for the marginal two-dimensional nonrelativistic point-interaction problem in Ref.~\cite{ErmanTurgut2013}; here it arises from the one-dimensional ultraviolet dispersion $\w(p)\sim|p|$.

The remaining optimization concerns the shape of the orbital.  The relevant
estimate is the one-dimensional fractional Gagliardo--Nirenberg inequality
\cite{FrankLenzmann2013},
\begin{equation}
 \int_{-\infty}^{\infty} \dd x\,|v(x)|^4
 \leq
 C_{\rm GN}\,
 \|v\|_2^2\,
 \langle v||\hat p||v\rangle ,
 \label{eq:fractionalGN}
\end{equation}
where $|\hat p|=(-\Delta)^{1/2}$ and $C_{\rm GN}$ denotes the optimal constant.
For normalized orbitals, $\|v\|_2=1$, this implies
\begin{equation}
 \frac{\displaystyle \int_{-\infty}^{\infty} \dd x\,|v(x)|^4}
 {\displaystyle\langle v||\hat p||v\rangle}
 \leq C_{\rm GN}.
\end{equation}
It is therefore natural to define
\begin{equation}
 C_{1/2}
 =
 \sup_{\|v\|_2=1}
 \frac{\displaystyle \int_{-\infty}^{\infty} \dd x\,|v(x)|^4}
 {\displaystyle\langle v||\hat p||v\rangle}
 =
 C_{\rm GN}.
 \label{eq:MFPOCsharp}
\end{equation}
The supremum is finite and is attained by an optimizer of the corresponding
fractional variational problem \cite{FrankLenzmann2013}.  Consequently, the
largest exponential coefficient allowed within the present asymptotic
mean-field approximation is
\begin{equation}
 c_*=\pi C_{1/2}.
 \label{eq:MFPOcstar}
\end{equation}
Determining $C_{1/2}$ numerically, or equivalently solving the associated nonlinear variational equation, would sharpen the large-$n$ estimate.

A completely explicit estimate is obtained from the normalized Lorentzian trial orbital
\begin{equation}
 u_a(x)=\sqrt{\frac{2a}{\pi}}\,\frac{1}{1+a^2x^2}.
 \label{eq:MFPOLorentzian}
\end{equation}
With the Fourier convention used in this paper,
\begin{equation}
 \widetilde u_a(p)=\sqrt{\frac{2\pi}{a}}\,e^{-|p|/a},
 \label{eq:MFPOLorentzianFT}
\end{equation}
so that
\begin{equation}
 \mathcal K[u_a]=\frac{a}{2},
 \qquad
 \mathcal Q[u_a]=\frac{5a}{4\pi},
 \qquad
 r=\frac{5}{2\pi}.
 \label{eq:MFPOLorentzianKQ}
\end{equation}
Hence $c=5/2$ and Eq.~\eqref{eq:MFPOkappatrial} gives the explicit large-$N$ trial estimate
\begin{equation}
 E_{\rm gr}^{\rm MF}
 \simeq
 -\frac{2\kappa_B}{5eN}
 \exp\!\left(\frac{5}{2}N\right),
 \qquad N=n-2\gg1.
 \label{eq:MFPOLorentzianEnergy}
\end{equation}
The coefficient $5/2$ is not universal; it is the value obtained from this simple trial shape.  Optimizing over $v$ replaces it by $c_*$ in Eq.~\eqref{eq:MFPOcstar}.

\paragraph{Massive case.}
The massive theory can be treated similarly, although the mass breaks the
exact scaling symmetry of the preceding calculation.  Let
\[
E=-\kappa,\qquad
\mathcal K_m[u]
=
\left\langle u\left|
\sqrt{\hat p^{\,2}+m^2}
\right|u\right\rangle,
\]
and define
\[
S=\kappa+N\mathcal K_m[u].
\]
In the deeply bound regime the heat-kernel integrals are dominated by short
times.  Using $[A_u(t)]^N\simeq e^{-Nt\mathcal K_m[u]}$ and
$B_\psi(t)\simeq K_{2t}(0)$, the pair-bubble contribution may be written in
terms of the exact two-body principal function as
\begin{equation}
 \mathcal F_n^{(0)}
 \simeq
 \Phi_2(0,-S)
 =
 \frac12F_m(E_B/2)
 +
 \frac{S}{2\pi\sqrt{S^2-4m^2}}
 \cosh^{-1}\!\left(\frac{S}{2m}\right),
 \qquad S>2m.
 \label{eq:MFPOmassiveF0}
\end{equation}
The leading $N^2$ term is again
\begin{equation}
 \mathcal F_n^{(2)}
 \simeq
 -\frac{N^2\mathcal Q[u]}{2S},
 \label{eq:MFPOmassiveF2}
\end{equation}
so that the massive mean-field equation becomes
\begin{equation}
 \pi F_m(E_B/2)
 +
 \frac{S}{\sqrt{S^2-4m^2}}
 \cosh^{-1}\!\left(\frac{S}{2m}\right)
 \simeq
 \frac{\pi N^2\mathcal Q[u]}{S}.
 \label{eq:MFPOmassiveImplicit}
\end{equation}
This equation retains the leading finite-mass dependence and may be used as
an implicit variational estimate of the binding energy.

For $S\gg m$,
\[
\Phi_2(0,-S)
=
\frac{1}{2\pi}
\ln\!\left(\frac{S}{\Lambda_B}\right)
+
O\!\left(
\frac{m^2}{S^2}\ln\frac{S}{m}
\right),
\]
where
\begin{equation}
 \Lambda_B
 =
 m\exp\!\left[-\pi F_m(E_B/2)\right].
 \label{eq:MFPOmassiveLambda}
\end{equation}
Thus the large-$N$ equation reduces to
\begin{equation}
 \ln\!\left(\frac{S}{\Lambda_B}\right)
 \simeq
 \frac{\pi N^2\mathcal Q[u]}{S}.
 \label{eq:MFPOmassiveLargeN}
\end{equation}
For a scaled orbital $u_a(x)=a^{1/2}v(ax)$,
\[
\mathcal K_m[u_a]
=
a\left\langle v\left|
\sqrt{\hat p^{\,2}+\frac{m^2}{a^2}}
\right|v\right\rangle
\simeq
a\mathcal K[v],
\qquad a\gg m,
\]
while $\mathcal Q[u_a]=a\mathcal Q[v]$.  The scale optimization therefore
coincides, to leading order, with that of the massless theory and gives
\begin{equation}
 E_{\rm gr}^{\rm MF}
 \sim
 -\frac{\Lambda_B}{e\,c_*N}
 \exp(c_*N),
 \qquad
 c_*=\pi C_{1/2},
 \qquad N=n-2\gg1.
 \label{eq:MFPOmassiveExp}
\end{equation}
In particular, the Lorentzian trial orbital used above gives
\begin{equation}
 E_{\rm gr}^{\rm MF}
 \simeq
 -\frac{2\Lambda_B}{5eN}
 \exp\!\left(\frac52N\right).
 \label{eq:MFPOmassiveLorentzian}
\end{equation}
The mass therefore changes the transmutation scale multiplying the
large-$N$ result, but not the leading ultraviolet exponential coefficient.
For binding energies comparable with $m$, the scale-invariant approximation
is no longer justified and Eq.~\eqref{eq:MFPOmassiveImplicit}, or the full
principal-operator mean-field functional, should be used instead.

\subsection{Brief comparison with an effective-coupling Hartree theory}
\label{subsec:MFcomparison}

A direct Hartree approximation to the regulated Hamiltonian is not
appropriate for removing the cutoff because the bare coupling satisfies
$\lambda(\epsilon)\to0$ as $\epsilon\to0^+$.  This does not make the
renormalized theory free; rather, a simple physical Hartree product does
not retain the short-distance contact-pair structure responsible for
dimensional transmutation.

For low-energy applications one may nevertheless replace the detailed
contact-pair dynamics by a finite coupling matched to the exact two-body
principal function at an off-shell energy $\mathcal E_*$,
\begin{equation}
 \frac{1}{g_{\rm R}(\mathcal E_*)}
 =
 \Phi_2(0,\mathcal E_*).
 \label{eq:MFPOgeff}
\end{equation}
This gives the effective stationary equation
\begin{equation}
 \left[
 \sqrt{-\partial_x^2+m^2}
 -(n-1)g_{\rm R}(\mathcal E_*)|\varphi(x)|^2
 \right]\varphi(x)
 =
 \mu\varphi(x),
 \label{eq:MFPOGP}
\end{equation}
which is a nonlocal semirelativistic analogue of the focusing
Gross--Pitaevskii equation.  It is simpler than the principal-operator
mean field, but it replaces the energy-dependent contact-pair dynamics
by one matched coupling and is therefore best regarded as a low-energy
effective approximation.

Restoring $\hbar$ and $c$ and taking the nonrelativistic limit,
$g_{\rm R}\to g_{\rm LL}$, gives the standard attractive
one-dimensional Hartree/Gross--Pitaevskii equation.  Thus this reduced
description provides a useful consistency check, whereas the
principal-operator mean field is the more direct approximation for the
renormalized binding problem.

\begin{remark}[Renormalization-group interpretation]
The logarithmic subtraction also gives a simple renormalization-group
interpretation.  Introducing an arbitrary energy scale $\mu$ and a
renormalized coupling $\lambda_R(\mu)$, independence of the principal
operator from $\mu$ yields, from
$K_{2t}(0)\sim(2\pi t)^{-1}$,
\begin{equation}
 \beta(\lambda_R)
 =
 \mu\frac{\dd\lambda_R}{\dd\mu}
 =
 -\frac{\lambda_R^2}{2\pi}.
\end{equation}
Thus the coupling decreases at high energies.  In the massless model,
where $E_B=-\kappa_B$,
\begin{equation}
 \frac{1}{\lambda_R(\mu)}
 =
 \frac{1}{2\pi}\ln\frac{\mu}{\kappa_B},
\end{equation}
so $\kappa_B$ is the renormalization-group invariant scale generated by
dimensional transmutation.  This is the one-dimensional Salpeter
counterpart of the mechanism discussed in
Ref.~\cite{ErmanTurgut2013}.
\end{remark}

\section{Existence and self-adjointness of the renormalized Hamiltonian}
\label{sec:selfadjoint}

For singular interactions it is often more convenient to construct the
resolvent first and reconstruct the Hamiltonian afterwards.  We follow the
direct resolvent strategy of Ref.~\cite{DoganErmanTurgut2012}.  The
argument is sectorwise, which is natural because particle number is
conserved.

\subsection{Pseudo-resolvent criterion}

Let $J(E)$ be a family of bounded operators on a Hilbert space satisfying
\begin{equation}
 J(E_1)-J(E_2)
 =
 (E_1-E_2)J(E_1)J(E_2).
 \label{eq:pseudoresidentity}
\end{equation}
Such a family is a pseudo-resolvent.  If it is defined on an unbounded
set and there exists a sequence $E_k$ with $|E_k|\to\infty$ such that
\begin{equation}
 -E_kJ(E_k)f\longrightarrow f
 \qquad\text{for every }f,
 \label{eq:pseudocriterion}
\end{equation}
then $J(E)$ is the resolvent of a unique densely defined closed operator
\cite{Pazy1983,DoganErmanTurgut2012}.  We use $E_k=-s_k$ with
$s_k\to\infty$.

\subsection{Bounded contact map after a free resolvent}

In the fixed $n$-boson sector define
\begin{equation}
 T_n(s)
 =
 \lim_{\epsilon\to0^+}
 b_{\epsilon,n}R_{0,n}(-s),
 \qquad
 R_{0,n}(-s)=(H_{0,n}+s)^{-1},
 \qquad s>0.
 \label{eq:Tns}
\end{equation}
Thus $T_n(s)$ maps
$\mathcal H_n$ into
$\mathcal H_{n-2}\otimes L^2(\mathbb R)$; no bare zero-cutoff contact
operator is being introduced.

Let
\begin{equation}
 \mathbf p=(p_1,\ldots,p_{n-2})
\end{equation}
denote the momenta of the $n-2$ ordinary bosons remaining after a pair
has been annihilated, and set
\begin{equation}
 \Omega_{n-2}(\mathbf p)
 =
 \sum_{j=1}^{n-2}\omega(p_j).
 \label{eq:Omegandef}
\end{equation}
Parametrize the annihilated pair by total and relative momenta,
\begin{equation}
 p_{n-1}=\frac P2+k,
 \qquad
 p_n=\frac P2-k.
\end{equation}
Then
\begin{align}
 &(T_n(s)\psi)(\mathbf p;P)
 \nonumber\\
 &\quad=
 c_n\int_{-\infty}^{\infty}\frac{\dd k}{2\pi}
 \frac{
 \psi(\mathbf p,P/2+k,P/2-k)}
 {
 s+\Omega_{n-2}(\mathbf p)
 +\omega(P/2+k)+\omega(P/2-k)
 },
 \label{eq:TactionD}
\end{align}
with
$c_n=\sqrt{n(n-1)/2}$.

Cauchy--Schwarz in $k$ gives
\begin{align}
 |(T_n(s)\psi)(\mathbf p;P)|^2
 \leq{}&
 c_n^2
 \int\frac{\dd k}{2\pi}
 \frac{1}{
 [s+\Omega_{n-2}
 +\omega(P/2+k)+\omega(P/2-k)]^2}
 \nonumber\\
 &\times
 \int\frac{\dd k}{2\pi}
 |\psi(\mathbf p,P/2+k,P/2-k)|^2 .
\end{align}
Since $\omega(q)\geq|q|$ and
$|P/2+k|+|P/2-k|\geq2|k|$,
\begin{equation}
 \int_{-\infty}^{\infty}\frac{\dd k}{2\pi}
 \frac{1}{(s+2|k|)^2}
 =
 \frac{1}{2\pi s}.
\end{equation}
Integration over $\mathbf p$ and $P$, followed by the unit-Jacobian
change from $(P,k)$ to the two pair momenta, therefore yields
\begin{equation}
 \|T_n(s)\|^2
 \leq
 \frac{n(n-1)}{4\pi s}.
 \label{eq:TboundD}
\end{equation}
The singular coincidence operation is thus bounded after one free
resolvent has acted.

\subsection{Invertibility of the principal operator at large negative energy}

For $E=-s$ write
\begin{equation}
 \Phi_n(-s)=K_n(-s)-U_n(-s),
 \label{eq:KUD}
\end{equation}
where
\begin{equation}
 K_n(-s)=\Phi_n^{(0)}(-s),
 \qquad
 U_n(-s)=-\Phi_n^{(1)}(-s)-\Phi_n^{(2)}(-s)\geq0.
\end{equation}
Appendix~\ref{app:estimates} proves that, for sufficiently large $s$,
\begin{equation}
 K_n(-s)
 \geq
 a_0\log(s/s_0)I
 \label{eq:KlowerD}
\end{equation}
with $a_0,s_0>0$, and that
\begin{equation}
 \widetilde U_n(-s)
 :=
 K_n(-s)^{-1/2}
 U_n(-s)
 K_n(-s)^{-1/2}
\end{equation}
satisfies
\begin{equation}
 0\leq\widetilde U_n(-s)
 \leq\eta_n(s)I,
 \qquad
 \eta_n(s)\to0
 \quad(s\to\infty).
 \label{eq:UtildeD}
\end{equation}
Choose $s$ so large that $\eta_n(s)<1/2$.  Then
\begin{equation}
 \Phi_n(-s)
 =
 K_n(-s)^{1/2}
 [I-\widetilde U_n(-s)]
 K_n(-s)^{1/2}
\end{equation}
is invertible and
\begin{equation}
 \|\Phi_n(-s)^{-1}\|
 \leq
 \frac{2}{a_0\log(s/s_0)}.
 \label{eq:PhiinverseD}
\end{equation}
The same factorization defines a self-adjoint principal operator at such
real spectral points by the standard form theorem
\cite{Kato1995}.

\subsection{Resolvent identity}

The difference identity for the principal operator is first obtained at
finite cutoff,
\[
 \Phi_{n,\epsilon}(E_1)-\Phi_{n,\epsilon}(E_2)
 =
 -b_{\epsilon,n}
 [R_{0,n}(E_1)-R_{0,n}(E_2)]
 b_{\epsilon,n}^\dagger.
\]
The counterterm drops out because it is independent of $E$.  Using the
free resolvent identity and then taking the cutoff-free limit in the
resolvent-dressed factors gives the bounded identity
\begin{equation}
 \Phi_n(E_1)-\Phi_n(E_2)
 =
 -(E_1-E_2)
 T_n(E_1)T_n(\overline E_2)^\dagger.
 \label{eq:PhiDifferenceD}
\end{equation}
Thus the singular zero-cutoff pair map never occurs by itself.

The renormalized sector resolvent is
\begin{equation}
 R_n(E)
 =
 R_{0,n}(E)
 +
 T_n(\overline E)^\dagger
 \Phi_n(E)^{-1}
 T_n(E).
 \label{eq:RnD}
\end{equation}
Combining Eq.~\eqref{eq:PhiDifferenceD} with
\[
 R_{0,n}(E_1)-R_{0,n}(E_2)
 =
 (E_1-E_2)R_{0,n}(E_1)R_{0,n}(E_2)
\]
and the elementary inverse identity
\[
 \Phi_n(E_1)^{-1}-\Phi_n(E_2)^{-1}
 =
 \Phi_n(E_1)^{-1}
 [\Phi_n(E_2)-\Phi_n(E_1)]
 \Phi_n(E_2)^{-1},
\]
a direct expansion gives
\begin{equation}
 R_n(E_1)-R_n(E_2)
 =
 (E_1-E_2)R_n(E_1)R_n(E_2).
 \label{eq:fullRID}
\end{equation}
Hence $R_n(E)$ is a pseudo-resolvent.

\subsection{Large-negative-energy limit}

At $E=-s$,
\begin{equation}
 R_n(-s)
 =
 R_{0,n}(-s)
 +
 T_n(s)^\dagger
 \Phi_n(-s)^{-1}
 T_n(s).
\end{equation}
For every $f\in\mathcal H_n$, the spectral theorem gives
\begin{equation}
 \|[sR_{0,n}(-s)-I]f\|
 \longrightarrow0.
\end{equation}
For the interaction term, Eqs.~\eqref{eq:TboundD} and
\eqref{eq:PhiinverseD} imply
\begin{align}
 s\left\|
 T_n(s)^\dagger\Phi_n(-s)^{-1}T_n(s)
 \right\|
 &\leq
 s\|T_n(s)\|^2\|\Phi_n(-s)^{-1}\|
 \nonumber\\
 &\leq
 \frac{n(n-1)}{4\pi}
 \frac{2}{a_0\log(s/s_0)}
 \longrightarrow0.
\end{align}
Hence
\begin{equation}
 sR_n(-s)f\longrightarrow f,
 \qquad s\to\infty.
 \label{eq:criterionprovedD}
\end{equation}
Equations~\eqref{eq:fullRID} and \eqref{eq:criterionprovedD} imply that
there exists a unique densely defined closed operator $H_n$ such that
\begin{equation}
 R_n(E)=(H_n-E)^{-1}.
\end{equation}

\subsection{Self-adjointness}

Choose one sufficiently large real $s$ for which the preceding estimates
hold.  The free resolvent is self-adjoint, the form construction gives a
self-adjoint $\Phi_n(-s)$, and therefore
\begin{equation}
 R_n(-s)
 =
 R_{0,n}(-s)
 +
 T_n(s)^\dagger\Phi_n(-s)^{-1}T_n(s)
\end{equation}
is bounded and self-adjoint.  Since
$R_n(-s)=(H_n+s)^{-1}$ is one-to-one with dense range, its inverse on its
range is self-adjoint.  Thus $H_n+s$ and hence $H_n$ are self-adjoint.

Finally, particle-number conservation permits the full operator to be
defined by
\begin{equation}
 H=\bigoplus_{n=0}^{\infty}H_n
\end{equation}
on
\begin{equation}
 \mathcal D(H)
 =
 \left\{
 \Psi=(\Psi_n):
 \Psi_n\in\mathcal D(H_n),\quad
 \sum_n\|H_n\Psi_n\|^2<\infty
 \right\}.
\end{equation}
This Hilbert-space direct sum is self-adjoint.  Self-adjointness is
logically distinct from stability as $n\to\infty$; an attractive
Fock-space Hamiltonian may be self-adjoint even if its spectrum is not
bounded from below uniformly in particle number.

\section*{Conclusions}

We have constructed a renormalized many-body theory for identical
one-dimensional bosons with spinless-Salpeter dispersion and attractive
pairwise contact interactions.  The linear ultraviolet behavior
$\omega(p)\sim|p|$ makes the contact interaction marginal and produces a
logarithmic divergence, in contrast with the ordinary one-dimensional
nonrelativistic Lieb--Liniger model.

Using an enlarged Fock space and a Schur-complement representation of the
resolvent, we isolate the divergent two-particle bubble and eliminate the
bare coupling in favor of the physical zero-total-momentum two-body
bound-state energy $E_B$.  The normal-ordered one- and zero-contraction
terms are ultraviolet finite, while the renormalized bubble grows
logarithmically at large negative energy.  Together with bounds on the
resolvent-dressed contact map, these estimates yield a self-adjoint
Hamiltonian in every fixed particle-number sector.

The two-body sector can be treated explicitly.  Its bound states are
determined by the zeros of the finite principal function
$\Phi_2(P,E)$, and the corresponding wave functions follow from residues
of the renormalized resolvent.  At zero total momentum the construction
reduces to the known one-dimensional Salpeter point interaction, while
in the massless theory the bound-pair dispersion can be expressed in
terms of the Lambert $W$ function.

We have also examined the nonrelativistic limit.  Writing
$E_B(c)=2mc^2-\varepsilon_B$ identifies the attractive Lieb--Liniger
coupling as
\[
 g_{\rm LL}=2\hbar\sqrt{\frac{\varepsilon_B}{m}}.
\]
At the operator level, Appendix~\ref{app:NRnorm} shows sectorwise
norm-resolvent convergence to the form-defined attractive Lieb--Liniger
Hamiltonian.  The trace-space description used there is only a convenient
representation of the usual coincidence interaction and does not enlarge
the physical Hilbert space of the Lieb--Liniger model.

Finally, we formulated a mean-field approximation directly for the
renormalized principal operator.  In the massless and deeply bound
large-particle-number regimes, the binding scale behaves exponentially,
\[
 |E_{\rm gr}^{\rm MF}|
 \sim
 \frac{\Lambda_B}{N}e^{cN},
 \qquad N=n-2,
\]
where the optimal coefficient is determined by a one-dimensional
fractional variational problem.  This behavior reflects the marginal
ultraviolet character of the interaction.  Natural extensions include a
detailed analysis of the three-body spectrum and scattering problem and
the exact large-particle-number stability properties of the renormalized
theory.

\appendix

\section{Normal-ordering calculation}\label{app:normal}

We give the field-theoretic algebra leading to Eqs.~\eqref{eq:Phi0}--\eqref{eq:Phi2op}.  Starting from Eq.~\eqref{eq:Phieps},
\begin{align}
 b_\ep R_0(E)b_\ep^\dagger
 =\frac12\int &\dd x\dd x'\dd x_1\dd x_2\dd x_1'\dd x_2'\,
 \chi^\dagger(x)
 K_\ep(x_1-x)K_\ep(x-x_2)
 K_\ep(x_1'-x')K_\ep(x'-x_2')
 \nonumber\\
 &\times
 \phi(x_1)\phi(x_2)
 (H_0-E)^{-1}
 \phi^\dagger(x_1')\phi^\dagger(x_2')
 \chi(x').
 \label{eq:app-start}
\end{align}
Using Eq.~\eqref{eq:pulltwo}, the creation operators are moved to the left of the free resolvent.  The remaining product of fields is
\begin{align}
 \phi(x_1)\phi(x_2)\phi^\dagger(y_1)\phi^\dagger(y_2)
 =&\,\phi^\dagger(y_1)\phi^\dagger(y_2)\phi(x_1)\phi(x_2)
 \nonumber\\
 &+\delta(x_2-y_1)\phi^\dagger(y_2)\phi(x_1)
 +\delta(x_2-y_2)\phi^\dagger(y_1)\phi(x_1)
 \nonumber\\
 &+\delta(x_1-y_1)\phi^\dagger(y_2)\phi(x_2)
 +\delta(x_1-y_2)\phi^\dagger(y_1)\phi(x_2)
 \nonumber\\
 &+\delta(x_1-y_1)\delta(x_2-y_2)
 +\delta(x_1-y_2)\delta(x_2-y_1).
 \label{eq:normalidentity}
\end{align}
The first line of Eq.~\eqref{eq:normalidentity} gives the zero-contraction term.  Applying the semigroup property to the original regulator kernels and the two propagation kernels gives, before removing the cutoff,
\begin{align}
 \mathcal U_{2,\ep}(E)=\frac12\int &\dd x\dd x'\dd x_1\dd x_2\dd x_1'\dd x_2'\,
 \chi^\dagger(x)\phi^\dagger(x_1')\phi^\dagger(x_2')
 \int_0^\infty\dd t\,
 K_{t+\ep}(x_1'-x')K_{t+\ep}(x'-x_2')
 \nonumber\\
 &\times K_{t+\ep}(x_1-x)K_{t+\ep}(x-x_2)
 \ee^{-t(H_0-E)}
 \phi(x_1)\phi(x_2)\chi(x').
 \label{eq:U2eps}
\end{align}
The four single contractions are equal after relabeling dummy variables, which gives
\begin{align}
 \mathcal U_{1,\ep}(E)=2\int &\dd x\dd x'\dd x_1\dd x_2\,
 \chi^\dagger(x)\phi^\dagger(x_1)
 \int_0^\infty\dd t\,
 K_{t+\ep}(x_1-x')K_{t+2\ep}(x'-x)K_{t+\ep}(x-x_2)
 \nonumber\\
 &\times\ee^{-t(H_0-E)}\phi(x_2)\chi(x').
 \label{eq:U1eps}
\end{align}
The two double contractions are also equal and cancel the overall factor $1/2$, giving
\begin{equation}
 \mathcal B_\ep(E)=
 \int\dd x\dd x'\,\chi^\dagger(x)
 \int_0^\infty\dd t\,
 K_{t+2\ep}(x-x')^2\ee^{-t(H_0-E)}\chi(x').
 \label{eq:Bubbleeps}
\end{equation}
At short time, Eq.~\eqref{eq:Ksquare} implies that only Eq.~\eqref{eq:Bubbleeps} contains a $\dd t/t$ singularity.  Combining it with the bare coupling \eqref{eq:barecoupling-heat} and taking $\ep\to0^+$ gives Eq.~\eqref{eq:Phi0}, while Eqs.~\eqref{eq:U1eps} and \eqref{eq:U2eps} reduce to Eqs.~\eqref{eq:Phi1} and \eqref{eq:Phi2op}.

\section{Ultraviolet finiteness of the normal-ordered terms}\label{app:estimates}

The purpose of this appendix is to explain, in a way that is useful for physics readers, why the subtraction of the two-body bubble is enough to remove the primitive ultraviolet divergence.  The essential point is simple: after normal ordering, the double contraction contains no remaining bosonic field operator that can act on a smooth many-body wave function, and therefore the two heat kernels collapse directly onto each other.  This produces the singular factor $K_{2t}(0)\sim(2\pi t)^{-1}$.  By contrast, the one- and zero-contraction pieces still contain annihilation operators.  Their heat kernels smear ordinary many-body wave functions before the limit $t\to0^+$ is taken, and the corresponding contact traces remain finite on a natural dense core.

We make this statement explicit below.  The estimates are not intended to be optimal; their role is to exhibit the ultraviolet mechanism and to provide the energy-weighted bounds used in Sec.~\ref{sec:selfadjoint}.

\subsection{A smooth dense core and what the heat kernels do}

Fix a total particle-number sector $n\ge2$.  The principal operator acts on
\begin{equation}
 \mathcal K_n=\Hh_{n-2}\otimes L^2(\R),
 \label{eq:angelspaceB}
\end{equation}
where the last factor describes the momentum, or equivalently the position, of the orthofermion.  We use the dense core
\begin{equation}
 \D_n=\mathcal S_{\rm sym}(\R^{n-2})\otimes\mathcal S(\R),
 \label{eq:densecore}
\end{equation}
with $\mathcal S$ denoting Schwartz space.  Thus all momentum-space wave functions, as well as all of their derivatives, decrease rapidly at infinity.

For a smooth function $f$,
\begin{equation}
 (K_t*f)(x)=\int\dd y\,K_t(x-y)f(y)
 \longrightarrow f(x),\qquad t\to0^+,
 \label{eq:approxidentity}
\end{equation}
and the convergence holds in every Schwartz seminorm.  Physically, the kernel does not create a singular function when it acts on a smooth state; it simply localizes the corresponding coordinate to a region of width of order $t$.

For example, consider a smooth $(n-2)$-boson wave function with one orthofermion,
\begin{equation}
 \Psi=\Psi(x_1,\ldots,x_{n-2};X_\chi).
 \label{eq:wavecore}
\end{equation}
A pair of heat kernels attached to two annihilation operators produces, schematically,
\begin{align}
 &\int\dd y_1\dd y_2\,
 K_t(y_1-X_\chi)K_t(y_2-X_\chi)
 \Psi(\ldots,y_1,y_2;X_\chi)
 \nonumber\\
 &\hspace{25mm}\xrightarrow[t\to0^+]{}
 \Psi(\ldots,X_\chi,X_\chi;X_\chi).
 \label{eq:contacttracepair}
\end{align}
Thus the short-time limit is simply a contact value of a smooth wave function.  In one dimension this value is finite on the core \eqref{eq:densecore}.  This observation is the main reason why the normal-ordered pieces behave differently from the bubble.

\subsection{Factorized form of the two finite pieces}

It is convenient to display the negative normal-ordered terms as positive quadratic forms.  Define
\begin{align}
 C_t&=\int\dd x\dd x_1\dd x_2\,
 K_t(x_1-x)K_t(x-x_2)
 \phi(x_1)\phi(x_2)\chi(x),
 \label{eq:Ct}\\
 D_{t,z}&=\int\dd x\dd x_1\,
 K_t(x_1-x)K_{t/2}(x-z)
 \phi(x_1)\chi(x).
 \label{eq:Dtz}
\end{align}
Up to harmless redistributions of the heat times by the semigroup property, Eqs.~\eqref{eq:Phi1} and \eqref{eq:Phi2op} can be written as
\begin{align}
 -\Phi^{(2)}(E)
 &=\frac12\int_0^\infty\dd t\,
 C_t^\dagger\ee^{-t(H_0-E)}C_t,
 \label{eq:Phi2factor}\\
 -\Phi^{(1)}(E)
 &=2\int_0^\infty\dd t\int\dd z\,
 D_{t,z}^\dagger\ee^{-t(H_0-E)}D_{t,z}.
 \label{eq:Phi1factor}
\end{align}
For real $E$ below the corresponding free threshold, the semigroup in these formulas is positive.  Hence both $-\Phi^{(1)}$ and $-\Phi^{(2)}$ are positive quadratic forms.  This factorization will also be useful in Sec.~\ref{sec:selfadjoint}.

On $\D_n$, the approximate-identity property gives the stronger
convergence statements
\begin{equation}
 \|C_t\Psi-C_0\Psi\|\longrightarrow0,
 \qquad t\to0^+,
 \label{eq:Ctlimit}
\end{equation}
and
\begin{equation}
 \int_{\mathbb R}\dd z\,
 \|(D_{t,z}-D_{0,z})\Psi\|^2
 \longrightarrow0,
 \qquad t\to0^+.
 \label{eq:Dtlimit}
\end{equation}
The symbols $C_0$ and $D_{0,z}$ are shorthand for coincidence traces on
the smooth core, not operators on the whole Hilbert space.  If
$N=n-2$ and
$\Psi_N(x_1,\ldots,x_N;X_\chi)$ is a one-orthofermion wave function, then
\begin{align}
 &(C_0\Psi_N)(x_1,\ldots,x_{N-2})
 \nonumber\\
 &\qquad=
 \sqrt{N(N-1)}
 \int_{\mathbb R}\dd x\,
 \Psi_N(x_1,\ldots,x_{N-2},x,x;x),
 \label{eq:C0action}
\end{align}
whereas
\begin{equation}
 (D_{0,z}\Psi_N)(x_1,\ldots,x_{N-1})
 =
 \sqrt N\,
 \Psi_N(x_1,\ldots,x_{N-1},z;z).
 \label{eq:D0action}
\end{equation}
These expressions are finite for $\Psi\in\D_n$.

Equations~\eqref{eq:Ctlimit} and \eqref{eq:Dtlimit} imply uniform
small-$t$ bounds.  For some $t_0>0$,
\begin{equation}
 \|C_t\Psi\|
 \leq
 \|C_0\Psi\|+1
 \equiv M_2(\Psi),
 \qquad 0<t<t_0,
 \label{eq:Cbound}
\end{equation}
and
\begin{equation}
 \int_{\mathbb R}\dd z\,\|D_{t,z}\Psi\|^2
 \leq M_1(\Psi),
 \qquad 0<t<t_0,
 \label{eq:Dbound}
\end{equation}
where one may take
\begin{equation}
 M_1(\Psi)
 =
 \left[
 1+
 \left(
 \int_{\mathbb R}\dd z\,\|D_{0,z}\Psi\|^2
 \right)^{1/2}
 \right]^2.
\end{equation}

If $E$ lies below the free threshold in this sector, the semigroup is locally bounded as $t\to0^+$.  Consequently,
\begin{align}
 \int_0^{t_0}\dd t\,
 \left|\langle\Psi,\Phi^{(2)}(E)\Psi\rangle\right|_t
 &\le \frac{t_0}{2}\,M_2(\Psi)^2
 \sup_{0<t<t_0}\|\ee^{-t(H_0-E)}\|,
 \label{eq:smallt2}\\
 \int_0^{t_0}\dd t\,
 \left|\langle\Psi,\Phi^{(1)}(E)\Psi\rangle\right|_t
 &\le 2t_0\,M_1(\Psi)
 \sup_{0<t<t_0}\|\ee^{-t(H_0-E)}\|.
 \label{eq:smallt1}
\end{align}
The right-hand sides are finite.  Therefore neither normal-ordered term contains a primitive short-time divergence.

The large-$t$ end is simpler.  After the annihilation operators in
$C_t$ or $D_{t,z}$ have acted, the semigroup propagates the corresponding
remaining finite-particle sector.  If that sector contains $r$ ordinary
bosons, its free threshold is $rm$ for $m>0$; in the massless case the
threshold is zero.  Taking $E$ below the relevant threshold gives
exponential decay at large $t$.  Hence the ultraviolet issue is entirely
the short-time endpoint analyzed above.

\subsection{Why the bubble is different}

The double contraction leaves no ordinary bosonic field operator.  There is therefore no smooth many-body wave function on which the two heat kernels can act separately.  They multiply directly and the spatial integration gives
\begin{equation}
 \int\dd r\,K_t(r)^2=K_{2t}(0)
 \sim\frac{1}{2\pi t},
 \qquad t\to0^+.
 \label{eq:bubbleonly}
\end{equation}
Thus the bubble contains
\begin{equation}
 \int_0^{t_0}\frac{\dd t}{t},
 \label{eq:logbubbleB}
\end{equation}
which is logarithmically divergent.  This is precisely the divergence removed by the physical subtraction in Eq.~\eqref{eq:barecoupling-heat}.

The distinction can therefore be summarized as follows:
\begin{equation}
 \begin{array}{ccl}
 \text{two contractions} &:& K_t^2\to K_{2t}(0)\sim t^{-1},\\[1mm]
 \text{one or zero contractions} &:& K_t*f\to f\quad\text{on smooth states}.
 \end{array}
 \label{eq:UVsummary}
\end{equation}
This is the operator version of the usual diagrammatic statement that only the two-particle bubble is primitively ultraviolet divergent.

\subsection{Momentum-space estimate needed for the existence proof}
\label{subsec:energyweighted}

For the existence and self-adjointness argument we need quantitative
control of the finite normal-ordered contribution at large negative
energy.  Put
\[
 E=-s,\qquad s>0,
\]
and write
\begin{equation}
 \Phi_n(-s)=K_n(-s)-U_n(-s),
 \label{eq:KUdecompositionB}
\end{equation}
where
\begin{equation}
 K_n(-s)=\Phi_n^{(0)}(-s),
 \qquad
 U_n(-s)=U_n^{(1)}(-s)+U_n^{(2)}(-s),
\end{equation}
with
\[
 U_n^{(1)}(-s)=-\Phi_n^{(1)}(-s),
 \qquad
 U_n^{(2)}(-s)=-\Phi_n^{(2)}(-s).
\]
The factorized forms above show that both are positive quadratic forms.

We use $\omega_p=\sqrt{p^2+m^2}$.  Up to relabeling dummy momenta, the
one-contraction term has the momentum-space form
\begin{align}
 U_n^{(1)}(-s)
 =2\int \frac{\dd p_1}{2\pi}\frac{\dd p_2}{2\pi}
       \frac{\dd q_2}{2\pi}\,
 &\chi^\dagger(p_1+p_2)a^\dagger(q_2)
 \nonumber\\
 &\times
 \frac{1}{
 H_0+s+\omega_{p_1}+\omega_{p_2}+\omega_{q_2}}
 a(p_2)\chi(p_1+q_2),
 \label{eq:U1momentumB}
\end{align}
whereas the zero-contraction term can be written
\begin{align}
 U_n^{(2)}(-s)
 =\frac12
 \int \frac{\dd p_1}{2\pi}\frac{\dd p_2}{2\pi}
      \frac{\dd q_1}{2\pi}\frac{\dd q_2}{2\pi}\,
 &\chi^\dagger(p_1+p_2)
 a^\dagger(q_1)a^\dagger(q_2)
 \nonumber\\
 &\times
 \frac{1}{
 H_0+s+\omega_{p_1}+\omega_{p_2}
 +\omega_{q_1}+\omega_{q_2}}
 \nonumber\\
 &\times
 a(p_1)a(p_2)\chi(q_1+q_2).
 \label{eq:U2momentumB}
\end{align}

Let $N=n-2$ denote the number of ordinary bosons in the
one-orthofermion sector.  Taking the quadratic form of
Eq.~\eqref{eq:U1momentumB}, using $H_0\ge0$ and
$\omega(p)\ge|p|$, and then applying Cauchy--Schwarz in the momentum
carried by the orthofermion gives
\begin{equation}
 \langle\Psi,U_n^{(1)}(-s)\Psi\rangle
 \le
 2\int_{\mathbb R^2}
 \frac{\dd p}{2\pi}\frac{\dd q}{2\pi}\,
 \frac{h_\Psi(p)h_\Psi(q)}
 {s+|p|+|q|},
 \label{eq:U1hB}
\end{equation}
where
\begin{equation}
 h_\Psi(p)^2
 =
 \int_{\mathbb R}\frac{\dd P}{2\pi}
 \|a(p)\chi(P)\Psi\|_{\mathcal H_{N-1}}^2.
 \label{eq:hPsiB}
\end{equation}
The integral form of Hilbert's inequality
\cite{HardyLittlewoodPolya1952} implies
\begin{equation}
 \int_{\mathbb R^2}
 \frac{|f(p)|\,|g(q)|}
 {a+|p|+|q|}
 \frac{\dd p}{2\pi}\frac{\dd q}{2\pi}
 \le C_{\rm H}\|f\|_2\|g\|_2,
 \qquad a\ge0,
 \label{eq:HilbertB}
\end{equation}
for a numerical constant $C_{\rm H}$.  Moreover, the number-operator
identity gives
\begin{equation}
 \int_{\mathbb R}\frac{\dd p}{2\pi}h_\Psi(p)^2
 =
 N\|\Psi\|^2.
 \label{eq:numberoneB}
\end{equation}
Hence
\begin{equation}
 0\le
 \langle\Psi,U_n^{(1)}(-s)\Psi\rangle
 \le C_1N\|\Psi\|^2,
 \label{eq:U1boundB}
\end{equation}
with $C_1$ independent of $s$.

The zero-contraction term is treated similarly.  Introduce total and
relative momenta
\[
 P=p_1+p_2,\qquad
 k=\frac{p_1-p_2}{2},
\]
and analogously $(Q,\ell)$ for the second pair.  Since
\begin{equation}
 \omega(P/2+k)+\omega(P/2-k)\ge2|k|,
\end{equation}
the energy denominator is bounded below by
$s+2|k|+2|\ell|$.  With
\begin{equation}
 G_{P,Q}(k)
 =
 \left\|
 a(P/2+k)a(P/2-k)\chi(Q)\Psi
 \right\|_{\mathcal H_{N-2}},
\end{equation}
Hilbert's inequality in $k,\ell$ followed by Cauchy--Schwarz in $P,Q$
gives
\begin{equation}
 \langle\Psi,U_n^{(2)}(-s)\Psi\rangle
 \le
 C_2
 \int
 \frac{\dd Q}{2\pi}\frac{\dd P}{2\pi}\frac{\dd k}{2\pi}
 G_{P,Q}(k)^2.
\end{equation}
The change of variables $(P,k)\leftrightarrow(p_1,p_2)$ has unit
Jacobian, and the two-particle number identity yields
\begin{equation}
 \int
 \frac{\dd Q}{2\pi}\frac{\dd P}{2\pi}\frac{\dd k}{2\pi}
 G_{P,Q}(k)^2
 =
 N(N-1)\|\Psi\|^2.
 \label{eq:numbertwoB}
\end{equation}
Consequently,
\begin{equation}
 0\le
 \langle\Psi,U_n^{(2)}(-s)\Psi\rangle
 \le
 C_2N(N-1)\|\Psi\|^2.
 \label{eq:U2boundB}
\end{equation}
Thus, for every fixed $n$,
\begin{equation}
 0\le U_n(-s)\le C_n I,
 \qquad
 C_n=C_1N+C_2N(N-1),
 \label{eq:UuniformB}
\end{equation}
as a quadratic-form inequality, uniformly at large negative energy.

It remains to compare this bounded finite part with the logarithmically
growing renormalized bubble.  In momentum representation $K_n(-s)$ is
multiplication by
\begin{align}
 \kappa_{n,s}(\mathbf p;P)
 =
 \int_{\mathbb R}\frac{\dd k}{2\pi}
 \Bigg[
 &\frac{1}{2\omega(k)-E_B}
 \nonumber\\
 &-
 \frac{1}{
 s+\Omega_{n-2}(\mathbf p)
 +\omega(P/2+k)+\omega(P/2-k)}
 \Bigg],
 \label{eq:kappans}
\end{align}
where
\[
 \Omega_{n-2}(\mathbf p)
 =
 \sum_{j=1}^{n-2}\omega(p_j).
\]
Convexity of $\omega$ gives
\[
 \omega(P/2+k)+\omega(P/2-k)\ge2\omega(k),
\]
and therefore
\begin{align}
 \kappa_{n,s}(\mathbf p;P)
 &\ge
 I(s)
 \nonumber\\
 &:=
 \int_{\mathbb R}\frac{\dd k}{2\pi}
 \left[
 \frac{1}{2\omega(k)-E_B}
 -
 \frac{1}{2\omega(k)+s}
 \right].
 \label{eq:IsB}
\end{align}
Let $k_0=m+|E_B|$ and take $s\ge8k_0$.  On
$k_0\le|k|\le s/8$,
\[
 2\omega(k)-E_B\le5|k|,
 \qquad
 2\omega(k)+s\le\frac{3s}{2},
 \qquad
 s+E_B\ge\frac{s}{2},
\]
so
\[
 \frac{s+E_B}{
 [2\omega(k)-E_B][2\omega(k)+s]}
 \ge\frac{1}{15|k|}.
\]
Restricting the integral to this interval gives
\begin{equation}
 I(s)
 \ge
 \frac{1}{15\pi}
 \log\left(\frac{s}{8k_0}\right).
\end{equation}
Hence there exist constants $a_0,s_0>0$, independent of the external
momenta, such that
\begin{equation}
 K_n(-s)
 \ge
 a_0\log(s/s_0)I
 \label{eq:KlogB}
\end{equation}
for sufficiently large $s$.  Positivity and functional calculus imply
\begin{equation}
 K_n(-s)^{-1}
 \le
 \frac{1}{a_0\log(s/s_0)}I.
\end{equation}
Combining this with Eq.~\eqref{eq:UuniformB},
\begin{equation}
 0\le
 K_n(-s)^{-1/2}U_n(-s)K_n(-s)^{-1/2}
 \le
 \eta_n(s)I,
\end{equation}
where
\begin{equation}
 \eta_n(s)
 \le
 \frac{C_n}{a_0\log(s/s_0)}
 \longrightarrow0.
 \label{eq:etaB}
\end{equation}
This is the relative estimate used in
Sec.~\ref{sec:selfadjoint}.  In particular, for sufficiently large $s$
the operator
$I-K_n^{-1/2}U_nK_n^{-1/2}$ is strictly positive and invertible.

\section{Norm-resolvent nonrelativistic limit}
\label{app:NRnorm}

In this appendix we study the nonrelativistic limit directly at the level
of the renormalized many-body resolvent.  We show, for every fixed
particle number $n$, that after subtraction of the rest energy the
semirelativistic Hamiltonian approaches the attractive Lieb--Liniger
Hamiltonian in the norm-resolvent sense.

The argument also identifies explicitly the limiting Lieb--Liniger
resolvent, so the convergence is established for the full renormalized
resolvent rather than only at the level of formal Hamiltonians or
bound-state conditions.

Throughout the intermediate estimates we set $\hbar=1$.  It will be
restored when the Lieb--Liniger coupling is identified.

We distinguish throughout between the momentum operator $\hat p$ and its
spectral variable $p$.  Thus $\tau_c(p)$ denotes a scalar dispersion
function, while $\tau_c(\hat p)$ denotes the corresponding one-particle
operator defined by functional calculus.  The quantities $p_j$, $P$, and
$k$ appearing in momentum-space kernels below are ordinary scalar
momenta.  Boldface symbols such as $\mathbf p$ and $\mathbf q$ denote
collections of such one-dimensional momentum variables, not spatial
vectors.

\subsection{Rest-energy-subtracted dispersion}

Define the scalar rest-energy-subtracted dispersion
\begin{equation}
 \tau_c(p)
 =
 \sqrt{m^2c^4+c^2p^2}-mc^2.
 \label{eq:tau-c-app}
\end{equation}
The nonrelativistic dispersion is $\tau_\infty(p)=\frac{p^2}{2m}$.
Rationalization gives $\tau_c(p)=\frac{p^2}{
m+\sqrt{m^2+p^2/c^2}}$. Hence $0\leq\tau_c(p)\leq\tau_\infty(p)$,
$\tau_c(p)\nearrow\tau_\infty(p)
\quad(c\to\infty)$.
A particularly useful exact identity is $\frac{p^2}{2m}
=\tau_c(p)+
\frac{\tau_c(p)^2}{2mc^2}$.
The corresponding one-particle operator is $\tau_c(\hat p)
=\sqrt{m^2c^4+c^2\hat p^{\,2}}-mc^2$, and the scalar identity above also holds by functional calculus:
$\frac{\hat p^{\,2}}{2m}
=\tau_c(\hat p)
+
\frac{\tau_c(\hat p)^2}{2mc^2}$.

Direct differentiation gives
$0\leq
 \tau_c''(p)
 =
 \frac{m^2c^6}{
 (m^2c^4+c^2p^2)^{3/2}}
 \leq\frac1m $. Therefore, for all $a,k\in\mathbb R$,
\begin{equation}
 0\leq
 \tau_c(k+a)+\tau_c(k-a)-2\tau_c(k)
 \leq
 \frac{a^2}{m}.
\end{equation}
In particular,
\[
 \tau_c(P/2+k)+\tau_c(P/2-k)\geq2\tau_c(k),
 \qquad
 \tau_c(P/2+k)+\tau_c(P/2-k)\geq2\tau_c(P/2).
\]
Averaging gives
\begin{equation}
 \tau_c(P/2+k)+\tau_c(P/2-k)
 \geq
 \tau_c(k)+\tau_c(P/2).
 \label{eq:tau-pair-average-app}
\end{equation}

The physical two-body subtraction energy is scaled as $E_B(c)=2mc^2-\varepsilon_B$, with $\varepsilon_B>0$, where $\varepsilon_B$ remains fixed as $c\to\infty$.

\subsection{Free resolvent}

In the fixed $n$-boson sector define
$K_{0,n}^{(c)}
 =
 H_{0,n}^{(c)}-nmc^2
 =
 \sum_{j=1}^{n}\tau_c(\hat p_j)$. The limiting nonrelativistic free Hamiltonian is
\begin{equation}
 H_{0,n}^{\rm NR}
 =
 \sum_{j=1}^{n}\frac{\hat p_j^{\,2}}{2m}.
\end{equation}
For $s>0$ we introduce $R_{0,n}^{(c)}(-s)
 =
 (K_{0,n}^{(c)}+s)^{-1}$ and $R_{0,n}^{\rm NR}(-s)
 =
 (H_{0,n}^{\rm NR}+s)^{-1}$.

To compare them, we pass to momentum representation.  For $\mathbf q=(q_1,\ldots,q_n)\in\mathbb R^n$, we define
$A_c(\mathbf q)
=\sum_{j=1}^{n}\tau_c(q_j)$ and
$A_\infty(\mathbf q)=
 \sum_{j=1}^{n}\frac{q_j^2}{2m}$. The exact dispersion identity gives
\begin{align}
 0\leq
 A_\infty(\mathbf q)-A_c(\mathbf q)
 =
 \frac{1}{2mc^2}
 \sum_{j=1}^{n}\tau_c(q_j)^2
 \leq
 \frac{A_c(\mathbf q)^2}{2mc^2}.
\end{align}
Consequently,
\begin{align}
 0
 &\leq
 \frac1{s+A_c(\mathbf q)}
 -
 \frac1{s+A_\infty(\mathbf q)}
 \nonumber\\
 &=
 \frac{
 A_\infty(\mathbf q)-A_c(\mathbf q)}
 {[s+A_c(\mathbf q)][s+A_\infty(\mathbf q)]}
 \nonumber\\
 &\leq
 \frac{1}{2mc^2}
 \frac{A_c(\mathbf q)^2}
 {[s+A_c(\mathbf q)][s+A_\infty(\mathbf q)]}.
\end{align}
Since $s>0$ and
$A_\infty(\mathbf q)\geq A_c(\mathbf q)\geq0$,
\begin{align}
 \frac{A_c(\mathbf q)^2}
 {[s+A_c(\mathbf q)][s+A_\infty(\mathbf q)]}
 &=
 \frac{A_c(\mathbf q)}
 {s+A_c(\mathbf q)}
 \frac{A_c(\mathbf q)}
 {s+A_\infty(\mathbf q)} \leq1.
\end{align}
Thus
\begin{equation}
 0
 \leq
 \frac1{s+A_c(\mathbf q)}
 -
 \frac1{s+A_\infty(\mathbf q)}
 \leq
 \frac1{2mc^2}.
\end{equation}
The two free resolvents are multiplication operators with respective
multipliers
$r_c(\mathbf q)=\frac1{s+A_c(\mathbf q)}$ and
$r_\infty(\mathbf q)=\frac1{s+A_\infty(\mathbf q)}$. Therefore
\begin{align}
 \left\|
 R_{0,n}^{(c)}(-s)
 -
 R_{0,n}^{\rm NR}(-s)
 \right\|_{\mathcal B(\mathcal H_n)}
 &=
 \operatorname*{ess\,sup}_{\mathbf q\in\mathbb R^n}
 |r_c(\mathbf q)-r_\infty(\mathbf q)|
 \nonumber\\
 &\leq
 \frac1{2mc^2},
\end{align}
where $\|A\|_{\mathcal B(\mathcal H_n)}=
 \sup_{\substack{\Psi\in\mathcal H_n\\ \|\Psi\|=1}}
 \|A\Psi\|$. Since the relevant Hilbert space and the type of norm will be clear from
the context, we shall henceforth suppress the subscript
$\mathcal B(\mathcal H_n)$ and write simply $\|A\|$ for the operator norm. Hence
\begin{equation}
 R_{0,n}^{(c)}(-s)
 \xrightarrow[c\to\infty]{\|\cdot\|}
 R_{0,n}^{\rm NR}(-s).
 \label{eq:free-norm-conv-app}
\end{equation}

\subsection{Resolvent-dressed contact map}

Let
\[
 T_{n,c}(s)
 =
 \lim_{\epsilon\to0^+}b_{\epsilon,n}R_{0,n}^{(c)}(-s),
\]
and denote its nonrelativistic counterpart by $T_{n,\infty}(s)$.
The symbol $b$ used below in intermediate formulas is only shorthand for
this coincidence operation after one free resolvent has acted; no bare
zero-cutoff pair-annihilation operator is required.

We now derive explicitly the momentum-space action of the
resolvent-dressed contact map $T_{n,c}(s)=
 bR_{0,n}^{(c)}(-s)$. Let
\begin{equation}
 |\Psi\rangle
 =
 \frac{1}{\sqrt{n!}}
 \int
 \prod_{\alpha=1}^{n}
 \frac{\dd q_\alpha}{2\pi}\,
 \Psi(q_1,\ldots,q_n)\,
 a^\dagger(q_1)\cdots a^\dagger(q_n)|0\rangle ,
 \label{eq:n-boson-state-C}
\end{equation}
where the momentum-space wave function
$\Psi(q_1,\ldots,q_n)$ is symmetric in all of its arguments.

The free resolvent
acts by multiplication:
\begin{align}
 &R_{0,n}^{(c)}(-s)|\Psi\rangle
 \nonumber\\
 &=
 \frac{1}{\sqrt{n!}}
 \int
 \prod_{\alpha=1}^{n}
 \frac{\dd q_\alpha}{2\pi}\,
 \frac{
 \Psi(q_1,\ldots,q_n)
 }{
 s+\displaystyle\sum_{\alpha=1}^{n}\tau_c(q_\alpha)
 }
 a^\dagger(q_1)\cdots a^\dagger(q_n)|0\rangle .
 \label{eq:free-res-action-C}
\end{align}
It is convenient to denote the corresponding wave function by
\begin{equation}
 \Psi_s^{(c)}(q_1,\ldots,q_n)
 =
 \frac{
 \Psi(q_1,\ldots,q_n)
 }{
 s+\displaystyle\sum_{\alpha=1}^{n}\tau_c(q_\alpha)
 }.
 \label{eq:Psi-s-C}
\end{equation}

We next act with two annihilation operators.  Using
\[
 [a(p),a^\dagger(q)]
 =
 2\pi\delta(p-q)
\]
and the symmetry of the $n$-boson wave function, one obtains
\begin{align}
 &a(q)a(q')R_{0,n}^{(c)}(-s)|\Psi\rangle
 \nonumber\\
 &=
 \frac{\sqrt{n(n-1)}}{\sqrt{(n-2)!}}
 \int
 \prod_{j=1}^{n-2}
 \frac{\dd p_j}{2\pi}\,
 \Psi_s^{(c)}
 (p_1,\ldots,p_{n-2},q,q')
 \nonumber\\
 &\hspace{25mm}\times
 a^\dagger(p_1)\cdots
 a^\dagger(p_{n-2})|0\rangle .
 \label{eq:two-ann-C}
\end{align}
The factor $\sqrt{n(n-1)}$ is the usual bosonic combinatorial factor
associated with removing two particles from an $n$-boson state.

For the pair which is annihilated, introduce its total momentum $P$
and relative momentum $k$ through
\begin{equation}
 q=\frac{P}{2}+k,
 \qquad
 q'=\frac{P}{2}-k.
 \label{eq:pair-param-C}
\end{equation}
Thus
\[
 q+q'=P.
\]
The remaining $n-2$ bosons have momenta
\begin{equation}
 \mathbf p=(p_1,\ldots,p_{n-2})
 \in\mathbb R^{n-2}.
 \label{eq:p-vector-C}
\end{equation}
Substituting Eq.~\eqref{eq:pair-param-C} into
Eq.~\eqref{eq:two-ann-C} gives
\begin{align}
 &a\!\left(\frac P2+k\right)
 a\!\left(\frac P2-k\right)
 R_{0,n}^{(c)}(-s)|\Psi\rangle
 \nonumber\\
 &=
 \frac{\sqrt{n(n-1)}}{\sqrt{(n-2)!}}
 \int
 \prod_{j=1}^{n-2}
 \frac{\dd p_j}{2\pi}\,
 \Psi_s^{(c)}
 \left(
 \mathbf p,\frac P2+k,\frac P2-k
 \right)
 \nonumber\\
 &\hspace{25mm}\times
 a^\dagger(p_1)\cdots
 a^\dagger(p_{n-2})|0\rangle .
 \label{eq:two-ann-pair-C}
\end{align}

Using Eq.~\eqref{eq:Psi-s-C}, the resolvent-dressed wave function appearing
here is
\begin{align}
 &\Psi_s^{(c)}
 \left(
 \mathbf p,\frac P2+k,\frac P2-k
 \right)
 \nonumber\\
 &=
 \frac{
 \Psi\left(
 \mathbf p,\frac P2+k,\frac P2-k
 \right)
 }{
 D_c(\mathbf p,P,k)
 },
 \label{eq:Psi-s-pair-C}
\end{align}
where
\begin{align}
 D_c(\mathbf p,P,k)
 ={}&
 s+
 \sum_{j=1}^{n-2}\tau_c(p_j)
 \nonumber\\
 &+
 \tau_c\!\left(\frac P2+k\right)
 +
 \tau_c\!\left(\frac P2-k\right).
 \label{eq:Dc-app}
\end{align}
The denominator contains the free energies of all $n$ bosons:
$n-2$ ordinary bosons with momenta $p_1,\ldots,p_{n-2}$ and the two
bosons in the annihilated pair with momenta $P/2+k$ and $P/2-k$.

The formal zero-cutoff pair-annihilation map is
\begin{equation}
 b_{\rm formal}
 =
 \frac{1}{\sqrt2}
 \int_{-\infty}^{\infty}\frac{\dd P}{2\pi}
 \int_{-\infty}^{\infty}\frac{\dd k}{2\pi}\,
 \chi^\dagger(P)
 a\!\left(\frac P2+k\right)
 a\!\left(\frac P2-k\right).
 \label{eq:bformal-C}
\end{equation}
The expression below is understood as the zero-cutoff limit of
$b_\epsilon R_{0,n}^{(c)}(-s)$ rather than as the action of
$b_{\rm formal}$ by itself.  Substituting
Eq.~\eqref{eq:two-ann-pair-C} into Eq.~\eqref{eq:bformal-C}, we obtain
\begin{align}
 &bR_{0,n}^{(c)}(-s)|\Psi\rangle
 \nonumber\\
 &=
 \frac{c_n}{\sqrt{(n-2)!}}
 \int_{-\infty}^{\infty}\frac{\dd P}{2\pi}
 \prod_{j=1}^{n-2}
 \frac{\dd p_j}{2\pi}\,
 \chi^\dagger(P)
 a^\dagger(p_1)\cdots
 a^\dagger(p_{n-2})|0\rangle
 \nonumber\\
 &\quad\times
 \int_{-\infty}^{\infty}\frac{\dd k}{2\pi}\,
 \frac{
 \Psi\left(
 \mathbf p,\frac P2+k,\frac P2-k
 \right)
 }{
 D_c(\mathbf p,P,k)
 },
 \label{eq:bR0-action-C}
\end{align}
where
\begin{equation}
 c_n
 =
 \frac{\sqrt{n(n-1)}}{\sqrt2}
 =
 \sqrt{\frac{n(n-1)}{2}}.
 \label{eq:cn-C}
\end{equation}

Therefore the momentum-space wave function of the resulting
one-orthofermion state is
\begin{align}
 &(T_{n,c}(s)\Psi)(\mathbf p;P)
 \nonumber\\
 &\qquad=
 c_n
 \int_{-\infty}^{\infty}\frac{\dd k}{2\pi}\,
 \frac{
 \Psi\left(
 \mathbf p,\frac P2+k,\frac P2-k
 \right)
 }{
 D_c(\mathbf p,P,k)
 }.
 \label{eq:T-action-app}
\end{align}

Here $\mathbf p$ has exactly $n-2$ components.  The variable $P$ is
written separately because it is the total momentum carried by the
annihilated pair, and hence by the orthofermion in the auxiliary sector.
Thus the output wave function naturally depends on
\[
 (\mathbf p;P)
 =
 (p_1,\ldots,p_{n-2};P).
\]

The limiting denominator is
\begin{equation}
 D_\infty(\mathbf p,P,k)
 =
 s+\sum_{j=1}^{n-2}\frac{p_j^2}{2m}
 +\frac{P^2}{4m}
 +\frac{k^2}{m}.
\end{equation}

Applying the exact identity for $\tau_c$ to every momentum of the
intermediate $n$-particle state gives
\begin{equation}
 0
 \leq
 D_c^{-1}-D_\infty^{-1}
 \leq
 \frac1{2mc^2}.
\end{equation}

To control the integral in $k$, divide it into
$|k|\leq\Lambda$ and $|k|>\Lambda$.  On the compact part,
\begin{equation}
 \int_{|k|\leq\Lambda}
 \frac{\mathrm dk}{2\pi}
 |D_c^{-1}-D_\infty^{-1}|^2
 \leq
 \frac{\Lambda}{4\pi m^2c^4}.
\end{equation}

For the tail, fix $c_0>0$ and take $c\geq c_0$.  Since
$\tau_c\geq\tau_{c_0}$, $D_c(\mathbf p,P,k)
 \geq
 s+2\tau_{c_0}(k)$. Furthermore,
$D_\infty(\mathbf p,P,k)
\geq
s+\frac{k^2}{m}
\geq
s+2\tau_{c_0}(k)$. Therefore
\begin{equation}
 |D_c^{-1}-D_\infty^{-1}|
 \leq
 \frac{2}{s+2\tau_{c_0}(k)}.
\end{equation}
Since
\[
 \tau_{c_0}(k)
 =
 \sqrt{m^2c_0^4+c_0^2k^2}-mc_0^2
 \sim c_0|k|,
 \qquad |k|\to\infty,
\]
there exists $\Lambda>0$ such that
\[
 \tau_{c_0}(k)\geq \frac{c_0}{2}|k|,
 \qquad |k|>\Lambda.
\]
Hence, for $|k|>\Lambda$,
\[
 \frac{4}{[s+2\tau_{c_0}(k)]^2}
 \leq
 \frac{4}{c_0^2k^2}.
\]
The right-hand side is integrable at infinity, so that
\begin{align}
 \int_{|k|>\Lambda}\frac{\dd k}{2\pi}
 \frac{4}{[s+2\tau_{c_0}(k)]^2}
 &\leq
 \frac{4}{2\pi c_0^2}
 \int_{|k|>\Lambda}\frac{\dd k}{k^2}
 \nonumber\\
 &=
 \frac{4}{\pi c_0^2\Lambda}
 \longrightarrow0
\end{align}
as $\Lambda\to\infty$.  Thus the high-momentum tail can be made
arbitrarily small uniformly for $c\geq c_0$.

Hence
\begin{equation}
 \sup_{\mathbf p,P}
 \int_{-\infty}^{\infty}\frac{\mathrm dk}{2\pi}
 |D_c^{-1}-D_\infty^{-1}|^2
 \longrightarrow0
\end{equation}
as $c\to\infty$. Cauchy--Schwarz inequality then gives
\begin{align}
 &|(T_{n,c}-T_{n,\infty})\Psi(\mathbf p;P)|^2
 \nonumber\\
 &\quad\leq
 c_n^2
 \int_{-\infty}^{\infty} \frac{\mathrm dk}{2\pi}
 |D_c^{-1}-D_\infty^{-1}|^2
 \int_{-\infty}^{\infty} \frac{\mathrm dk}{2\pi}
 \left|
 \Psi\!\left(
 \mathbf p,P/2+k,P/2-k
 \right)
 \right|^2 .
\end{align}
After integration over $\mathbf p$ and $P$,
\begin{equation}
 \|T_{n,c}(s)-T_{n,\infty}(s)\|
 \longrightarrow0.
 \label{eq:T-norm-conv-app}
\end{equation}

The norm convergence in Eq.~\eqref{eq:T-norm-conv-app} is needed because the resolvent of the
interacting problem contains the contact map through the combination
\[
 T_{n,c}(s)^\dagger
 \Phi_{n,c}(-s)^{-1}
 T_{n,c}(s).
\]
Thus convergence of the free resolvent alone is not sufficient to establish
the nonrelativistic limit of the interacting theory.  Equation~\eqref{eq:T-norm-conv-app}
shows that the map connecting the $n$-boson sector with the auxiliary
one-orthofermion sector converges uniformly in operator norm to its
nonrelativistic counterpart.  Together with the convergence estimates for
the principal operator derived below, this will imply convergence of the
complete interaction correction to the resolvent and hence, ultimately,
the norm-resolvent convergence of the renormalized semirelativistic
Hamiltonian to the attractive Lieb--Liniger Hamiltonian.

\subsection{Renormalized bubble and identification of the limiting multiplier}

At the shifted energy
\begin{equation}
 E=nmc^2-s
\end{equation}
the renormalized bubble part is a multiplication operator on the
semirelativistic auxiliary contact space:
\begin{equation}
 (K_{n,c}(-s)\Psi)(\mathbf p;P)
 =
 \kappa_{n,c,s}(\mathbf p;P)\Psi(\mathbf p;P).
 \label{eq:Kmult-app}
\end{equation}
Its multiplier is
\begin{align}
 \kappa_{n,c,s}(\mathbf p;P)
 =
 \int_{-\infty}^{\infty}\frac{\mathrm dk}{2\pi}
 \Bigg[
 &\frac{1}{2\tau_c(k)+\varepsilon_B}
 \nonumber\\
 &-
 \frac{1}{
 s+\Omega_{n-2}^{(c)}(\mathbf p)
 +\tau_c(P/2+k)
 +\tau_c(P/2-k)}
 \Bigg],
\label{eq:kappa-c-app}
\end{align}
where $\Omega_{n-2}^{(c)}(\mathbf p)
 = \sum_{j=1}^{n-2}\tau_c(p_j)$. The limiting multiplier is
\begin{align}
 \kappa_{n,\infty,s}(\mathbf p;P)
 =
 \int_{-\infty}^{\infty}\frac{\mathrm dk}{2\pi}
 \Bigg[
 &\frac{1}{k^2/m+\varepsilon_B}
 \nonumber\\
 &-
 \frac{1}{
 s+\Omega_{n-2}^{\rm NR}(\mathbf p)
 +P^2/(4m)+k^2/m}
 \Bigg],
\label{eq:kappa-inf-app}
\end{align}
where $\Omega_{n-2}^{\rm NR}(\mathbf p)
 =
 \sum_{j=1}^{n-2}\frac{p_j^2}{2m}$. Using
\begin{equation}
 \int_{-\infty}^{\infty}\frac{\mathrm dk}{2\pi}
 \frac{1}{k^2/m+A}
 =
 \frac{\sqrt m}{2\sqrt A},
 \qquad A>0,
\end{equation}
we obtain
\begin{equation}
 \kappa_{n,\infty,s}(\mathbf p;P)
 =
 \frac{\sqrt m}{2}
 \left[
 \frac1{\sqrt{\varepsilon_B}}
 -
 \frac{1}{
 \sqrt{
 s+\Omega_{n-2}^{\rm NR}(\mathbf p)+P^2/(4m)}}
 \right].
 \label{eq:kappa-inf-explicit-app}
\end{equation}

For $s>\varepsilon_B$,
\begin{align}
 K_{n,c}(-s)
 \geq 
 \int_{-\infty}^{\infty}\frac{\mathrm dk}{2\pi}
 \left[
 \frac{1}{2\tau_c(k)+\varepsilon_B}
 -
 \frac{1}{2\tau_c(k)+s}
 \right]I
\geq
 \kappa_s I.
 \label{eq:K-lower-app}
\end{align}
where
\begin{equation}
 \kappa_s
 =
 \frac{\sqrt m}{2}
 \left(
 \frac1{\sqrt{\varepsilon_B}}
 -
 \frac1{\sqrt s}
 \right)>0.
 \label{eq:kappa-s-app}
\end{equation}
The second inequality follows because
$\tau_c(k)\leq k^2/(2m)$ and the difference of the two reciprocal
functions is decreasing in $\tau_c(k)$.

Hence
\begin{equation}
 \sup_{c\geq c_0}
 \|K_{n,c}(-s)^{-1/2}\|
 \leq
 \kappa_s^{-1/2}.
 \label{eq:Kinv-uniform-app}
\end{equation}
The same lower bound holds for $K_{n,\infty}(-s)$.

For every fixed $\Lambda$,
\begin{equation}
 \sup_{\substack{|p_j|\leq\Lambda\\|P|\leq\Lambda}}
 |\kappa_{n,c,s}(\mathbf p;P)
 -
 \kappa_{n,\infty,s}(\mathbf p;P)|
 \longrightarrow0.
 \label{eq:kappa-local-app}
\end{equation}
Indeed, on a bounded interval of the internal momentum $k$ all dispersions
converge uniformly.  For large $|k|$, the subtraction can be estimated
using
\begin{equation}
 0\leq
 \tau_c(P/2+k)+\tau_c(P/2-k)-2\tau_c(k)
 \leq
 \frac{P^2}{4m}.
\end{equation}
For bounded external momenta the subtracted integrand is consequently
dominated by
\begin{equation}
 \frac{C_{\Lambda,s,\varepsilon_B}}
 {[1+\tau_{c_0}(k)]^2},
\end{equation}
which is integrable.  Dominated convergence then proves the local uniform
limit.

\subsection{The attractive Lieb--Liniger resolvent}

We now identify the operator obtained from the limiting
nonrelativistic resolvent.  No enlargement of the bosonic Hilbert space
is required for the one-dimensional Lieb--Liniger model.  The contact
interaction can be defined directly on the original $n$-boson Hilbert
space through a closed, semibounded quadratic form.

Let
\begin{equation}
 \mathcal H_n=L^2_{\rm sym}(\mathbb R^n)
\end{equation}
and
\begin{equation}
 H_{0,n}^{\rm NR}
 =
 \sum_{j=1}^{n}\frac{\hat p_j^{\,2}}{2m}.
\end{equation}
The Lieb--Liniger model was introduced in
Ref.~\cite{LiebLiniger1963}, while the attractive many-body problem and
its bound states were studied in Ref.~\cite{McGuire1964}.  A rigorous
construction of one-dimensional many-particle contact interactions,
together with a generalized Krein-type resolvent representation, is
given in Ref.~\cite{GriesemerHofackerLinden2020}; see also
Ref.~\cite{ProlhacSpohn2011}.

For $\Psi\in H^1_{\rm sym}(\mathbb R^n)$, the Sobolev trace theorem
\cite{Evans2010,AdamsFournier2003} allows restriction of $\Psi$ to a
particle-coincidence hypersurface.  Introduce the trace map
\begin{equation}
 \mathcal T_n:
 H^1_{\rm sym}(\mathbb R^n)
 \longrightarrow
 \mathcal G_n,
 \qquad
 \mathcal G_n
 =
 L^2_{\rm sym}(\mathbb R^{n-2})\otimes L^2(\mathbb R),
 \label{eq:LLtrace-app}
\end{equation}
by
\begin{align}
 &(\mathcal T_n\Psi)(x_1,\ldots,x_{n-2};X)
 \nonumber\\
 &\qquad=
 c_n\,
 \Psi(x_1,\ldots,x_{n-2},X,X),
 \qquad
 c_n=\sqrt{\frac{n(n-1)}{2}}.
 \label{eq:LLtraceaction-app}
\end{align}
Here ``trace'' means restriction of a sufficiently regular function to a
lower-dimensional set; it is unrelated to the trace of an operator.  A
generic $L^2$ wave function cannot be evaluated on the measure-zero set
$x_i=x_j$, whereas the $H^1$ regularity is sufficient to define this
restriction as an $L^2$ function.

The space $\mathcal G_n$ is only the codomain of the Sobolev trace map.
It is not an enlarged physical Hilbert space and introduces no
orthofermionic degree of freedom into the Lieb--Liniger model.  Its
Hilbert-space realization happens to coincide with that of the
semirelativistic one-orthofermion sector, which is useful only for the
comparison below.

By bosonic symmetry,
\begin{align}
 \|\mathcal T_n\Psi\|_{\mathcal G_n}^2
 &=
 \sum_{1\leq i<j\leq n}
 \int_{\mathbb R^n}\dd^n x\,
 \delta(x_i-x_j)|\Psi(x_1,\ldots,x_n)|^2 .
 \label{eq:LLtracenorm-app}
\end{align}
The attractive Lieb--Liniger quadratic form is therefore
\begin{align}
 \mathfrak q_n^{\rm LL}[\Psi]
 ={}&
 \langle\Psi,H_{0,n}^{\rm NR}\Psi\rangle
 -
 g_{\rm LL}\|\mathcal T_n\Psi\|_{\mathcal G_n}^2,
 \nonumber\\
 &\Psi\in H^1_{\rm sym}(\mathbb R^n),
 \qquad g_{\rm LL}>0.
 \label{eq:LL-form-app}
\end{align}
The coincidence traces are infinitesimally form bounded with respect to
the free kinetic-energy form: for every $\varepsilon>0$ there exists
$C_{\varepsilon,n}<\infty$ such that
\begin{equation}
 \sum_{i<j}\|\gamma_{ij}\Psi\|^2
 \le
 \varepsilon
 \langle\Psi,H_{0,n}^{\rm NR}\Psi\rangle
 +
 C_{\varepsilon,n}\|\Psi\|^2.
 \label{eq:LLtraceformbound-app}
\end{equation}
A direct treatment of these coincidence traces for the one-dimensional
many-body contact problem is given in
Ref.~\cite{GriesemerHofackerLinden2020}.  It follows from the standard
quadratic-form perturbation theorem and the representation theorem for
closed semibounded forms \cite{Kato1995} that
$\mathfrak q_n^{\rm LL}$ determines a unique lower-semibounded
self-adjoint operator, denoted by $H_n^{\rm LL}$.  The formal expression
\begin{equation}
 H_n^{\rm LL}
 =
 H_{0,n}^{\rm NR}
 -
 g_{\rm LL}
 \sum_{i<j}\delta(x_i-x_j)
 \label{eq:LLformal-app}
\end{equation}
is understood as shorthand for this form-defined operator.

The trace notation permits the resolvent to be factorized in a form
directly comparable with the semirelativistic one.  Let
\[
 R_{0,n}^{\rm NR}(-s)=(H_{0,n}^{\rm NR}+s)^{-1}.
\]
For $-s\in\rho(H_n^{\rm LL})$, the quadratic-form resolvent equation may
be written formally as
\begin{equation}
 (H_{0,n}^{\rm NR}+s)\Psi
 -
 g_{\rm LL}\mathcal T_n^\dagger\mathcal T_n\Psi
 =
 F,
\end{equation}
where $\mathcal T_n^\dagger\mathcal T_n$ is only shorthand for the
contact form.  Introducing
$\eta=\mathcal T_n\Psi$ and
\begin{equation}
 \Gamma_n(s)
 =
 \mathcal T_nR_{0,n}^{\rm NR}(-s)\mathcal T_n^\dagger,
 \label{eq:GammaLL-app}
\end{equation}
one obtains
\[
 [I-g_{\rm LL}\Gamma_n(s)]\eta
 =
 \mathcal T_nR_{0,n}^{\rm NR}(-s)F.
\]
Whenever the bracket is invertible,
\begin{align}
 (H_n^{\rm LL}+s)^{-1}
 ={}&
 R_{0,n}^{\rm NR}(-s)
 \nonumber\\
 &+
 g_{\rm LL}
 R_{0,n}^{\rm NR}(-s)\mathcal T_n^\dagger
 [I-g_{\rm LL}\Gamma_n(s)]^{-1}
 \mathcal T_nR_{0,n}^{\rm NR}(-s).
\end{align}

For comparison with the semirelativistic notation define
\begin{equation}
 \Phi_n^{\rm LL}(-s)
 =
 \frac{1}{g_{\rm LL}}I-\Gamma_n(s).
 \label{eq:PhiLL-app}
\end{equation}
This is merely a convenient trace-space rewriting; the
Lieb--Liniger model itself does not require a principal-operator or
enlarged-Fock-space construction.  Then
\begin{align}
 (H_n^{\rm LL}+s)^{-1}
 ={}&
 R_{0,n}^{\rm NR}(-s)
 \nonumber\\
 &+
 R_{0,n}^{\rm NR}(-s)\mathcal T_n^\dagger
 [\Phi_n^{\rm LL}(-s)]^{-1}
 \mathcal T_nR_{0,n}^{\rm NR}(-s).
 \label{eq:LL-Krein-app}
\end{align}
With
\begin{equation}
 T_{n,\infty}(s)
 =
 \mathcal T_nR_{0,n}^{\rm NR}(-s),
 \label{eq:Tninf-app}
\end{equation}
this becomes
\begin{align}
 (H_n^{\rm LL}+s)^{-1}
 =
 R_{0,n}^{\rm NR}(-s)
 +
 T_{n,\infty}(s)^\dagger
 [\Phi_n^{\rm LL}(-s)]^{-1}
 T_{n,\infty}(s).
 \label{eq:LL-res-T-app}
\end{align}
The nonrelativistic contact loop is ultraviolet finite because its
relative-momentum denominator behaves as $k^{-2}$ at large $|k|$; no
ultraviolet coupling-constant renormalization is required.

\subsection{Matching the coupling to the binding energy}

For $n=2$, the trace-space operator is diagonal in the total pair
momentum $P$.  At energy $-s$,
\begin{equation}
 \Phi_2^{\rm LL}(P,-s)
 =
 \frac1{g_{\rm LL}}
 -
 \int_{-\infty}^{\infty}\frac{\dd k}{2\pi}
 \frac1{s+P^2/(4m)+k^2/m}.
\end{equation}
Using
\[
 \int_{-\infty}^{\infty}\frac{\dd k}{2\pi}
 \frac1{k^2/m+A}
 =
 \frac{\sqrt m}{2\sqrt A},
 \qquad A>0,
\]
we obtain
\begin{equation}
 \Phi_2^{\rm LL}(P,-s)
 =
 \frac1{g_{\rm LL}}
 -
 \frac{\sqrt m}{
 2\sqrt{s+P^2/(4m)}}.
\end{equation}
Restoring $\hbar$, the zero-total-momentum dimer condition with binding
energy $\varepsilon_B$ is
\begin{equation}
 \frac1{g_{\rm LL}}
 =
 \int_{-\infty}^{\infty}\frac{\dd k}{2\pi\hbar}
 \frac1{k^2/m+\varepsilon_B}
 =
 \frac{\sqrt m}{2\hbar\sqrt{\varepsilon_B}}.
\end{equation}
Hence
\begin{equation}
 g_{\rm LL}
 =
 2\hbar\sqrt{\frac{\varepsilon_B}{m}}.
 \label{eq:gLL-app}
\end{equation}
For general total momentum,
\begin{equation}
 \mathcal E_B(P)
 =
 \frac{P^2}{4m}-\varepsilon_B,
\end{equation}
the Galilean dispersion of a bound pair of total mass $2m$.

\subsection{Uniform momentum-space control and convergence of the finite terms}
\label{app:NR-uniform}

The convergence
$\tau_c(p)\to p^2/(2m)$ is uniform on bounded momentum intervals but not
on the whole real line.  We therefore separate bounded external momenta
from the high-momentum region.  The Hilbert space used for this
comparison is
\begin{equation}
 \mathcal G_n
 =
 L^2_{\rm sym}(\mathbb R^{n-2})\otimes L^2(\mathbb R).
\end{equation}
For the semirelativistic theory this is the Hilbert-space realization of
the one-orthofermion sector; for the Lieb--Liniger theory it is only the
codomain of the Sobolev trace map.  Identifying these two copies of
$\mathcal G_n$ is an analytical device for the limiting argument.

\subsubsection*{Momentum-space localization}

For $\Lambda>0$ let $\mathsf P_\Lambda$ be the orthogonal projection
defined by
\begin{align}
 (\mathsf P_\Lambda\Xi)(\mathbf p;P)
 =
 \mathbf 1_{\{
 |p_j|\leq\Lambda\ {\rm for\ all}\ j,\,
 |P|\leq\Lambda
 \}}
 \Xi(\mathbf p;P),
 \label{eq:Plambda-app}
\end{align}
and let
\begin{equation}
 \mathsf Q_\Lambda=I-\mathsf P_\Lambda.
\end{equation}
These projections occur only in the proof and are not additional
ultraviolet regulators.

Fix $c_0>0$.  For $c\geq c_0$, monotonicity of $\tau_c$ in $c$ and
Eq.~\eqref{eq:tau-pair-average-app} give
\begin{equation}
 D_c(\mathbf p,P,k)
 \geq
 A_{c_0}(\mathbf p,P)+\tau_{c_0}(k),
\end{equation}
where
\begin{equation}
 A_{c_0}(\mathbf p,P)
 =
 s+\sum_{j=1}^{n-2}\tau_{c_0}(p_j)
 +\tau_{c_0}(P/2).
\end{equation}
Define
\begin{equation}
 J_{c_0}(A)
 =
 \int_{-\infty}^{\infty}\frac{\dd k}{2\pi}
 \frac{1}{[A+\tau_{c_0}(k)]^2}.
 \label{eq:J-app}
\end{equation}
Because $\tau_{c_0}(k)\sim c_0|k|$, this integral is finite.  It is
decreasing in $A$ and dominated convergence gives
\begin{equation}
 J_{c_0}(A)\longrightarrow0,
 \qquad A\to\infty.
\end{equation}

On the support of $\mathsf Q_\Lambda$, at least one
$|p_j|>\Lambda$ or $|P|>\Lambda$, and therefore
\begin{equation}
 A_{c_0}(\mathbf p,P)
 \geq
 s+\tau_{c_0}(\Lambda/2).
\end{equation}
Applying Cauchy--Schwarz to Eq.~\eqref{eq:T-action-app} and then using
the unit-Jacobian transformation from $(P,k)$ to the two pair momenta
gives
\begin{equation}
 \|\mathsf Q_\Lambda T_{n,c}(s)\|^2
 \leq
 c_n^2
 J_{c_0}\!\left(
 s+\tau_{c_0}(\Lambda/2)
 \right),
 \qquad c\geq c_0.
 \label{eq:QT-high-app}
\end{equation}
Consequently,
\begin{equation}
 \lim_{\Lambda\to\infty}
 \sup_{c\geq c_0}
 \|\mathsf Q_\Lambda T_{n,c}(s)\|
 =
 0.
 \label{eq:QT-tight-app}
\end{equation}
The same argument applies to $T_{n,\infty}(s)$.

\subsubsection*{Finite contact-space contribution}

Write
\begin{equation}
 \Phi_{n,c}(-s)
 =
 K_{n,c}(-s)-U_{n,c}(-s),
 \label{eq:Phi-KU-app}
\end{equation}
where $K_{n,c}$ is the renormalized diagonal pair-bubble contribution
and $U_{n,c}\geq0$ is the sum of the remaining one- and
zero-contraction terms.  In the limiting Lieb--Liniger trace-space
representation, the same-pair part of
$\Gamma_n(s)$ gives the multiplier $K_{n,\infty}$ in
Eq.~\eqref{eq:kappa-inf-explicit-app}; the contributions in which the
two traces involve different particle pairs give $U_{n,\infty}$.  Thus
\begin{equation}
 \Phi_n^{\rm LL}(-s)
 =
 K_{n,\infty}(-s)-U_{n,\infty}(-s).
 \label{eq:PhiLL-KU-app}
\end{equation}
This decomposition of the Lieb--Liniger trace-space operator is purely
algebraic and does not represent a renormalization procedure.

For $s>\varepsilon_B$, Eq.~\eqref{eq:K-lower-app} gives the common
positive lower bound
\begin{equation}
 K_{n,c}(-s)\geq\kappa_s I,
 \qquad
 K_{n,\infty}(-s)\geq\kappa_s I,
 \label{eq:K-common-lower-app}
\end{equation}
and hence
\begin{equation}
 \sup_{c\geq c_0}\|K_{n,c}(-s)^{-1/2}\|
 \leq\kappa_s^{-1/2},
 \qquad
 \|K_{n,\infty}(-s)^{-1/2}\|
 \leq\kappa_s^{-1/2}.
\end{equation}
Set
\begin{equation}
 \widetilde U_{n,c}(s)
 =
 K_{n,c}(-s)^{-1/2}
 U_{n,c}(-s)
 K_{n,c}(-s)^{-1/2},
\end{equation}
and define $\widetilde U_{n,\infty}(s)$ analogously.

We now record the finite-part estimate used below.

\paragraph{Finite-part convergence lemma.}
For fixed $n$ and fixed $s>\varepsilon_B$ for which the limiting
trace-space operator is invertible,
\begin{equation}
 \|\widetilde U_{n,c}(s)-\widetilde U_{n,\infty}(s)\|
 \longrightarrow0,
 \qquad c\to\infty.
 \label{eq:Utilde-conv-app}
\end{equation}

\emph{Proof.}
In a fixed-$n$ sector there are only finitely many particle-pair
routings.  They fall into the same two classes displayed explicitly in
Eqs.~\eqref{eq:U1momentumB} and \eqref{eq:U2momentumB}: two contact
traces may share one ordinary boson, or they may involve disjoint
ordinary bosons.  Restoring $c$ replaces each free energy in those
kernels by the corresponding $\tau_c$, while the nonrelativistic kernels
are obtained by replacing $\tau_c(p)$ with $p^2/(2m)$.

On the range of $\mathsf P_\Lambda$, the external momenta are bounded.
The dispersions and the bubble multipliers then converge uniformly, and
the internal integrations are dominated by the squared-denominator
bound
\begin{equation}
 \int\frac{\dd k}{2\pi}
 \frac{1}{[A+\tau_c(k)]^2}
 \leq J_{c_0}(A),
 \qquad c\geq c_0.
 \label{eq:finite-basic-app}
\end{equation}
For the one-shared-particle routing, Cauchy--Schwarz followed by
Hilbert's integral inequality \eqref{eq:HilbertB} provides an
operator-norm majorant independent of $c$.  For the disjoint-pair
routing the same argument is applied to each relative momentum and is
then closed with the fixed-particle number identity
\eqref{eq:numbertwoB}.  Dominated convergence therefore yields
\begin{equation}
 \left\|
 \mathsf P_\Lambda
 [\widetilde U_{n,c}(s)-\widetilde U_{n,\infty}(s)]
 \mathsf P_\Lambda
 \right\|
 \longrightarrow0
 \label{eq:Utilde-local-app}
\end{equation}
for every fixed $\Lambda$.

For the complementary region, at least one external free energy is
bounded below by
$s+\tau_{c_0}(\Lambda/2)$.  Repeating the same Cauchy--Schwarz/Hilbert
estimates therefore extracts at least one factor
\[
 J_{c_0}\!\left(
 s+\tau_{c_0}(\Lambda/2)
 \right)^{1/2}.
\]
The remaining integrations are bounded by the number identities
\eqref{eq:numberoneB} and \eqref{eq:numbertwoB}.  Since there are only
finitely many routings, there is a constant $C_{n,s,c_0}$ such that
\begin{align}
 \sup_{c\geq c_0}
 \|\mathsf Q_\Lambda\widetilde U_{n,c}(s)\|
 &\leq
 C_{n,s,c_0}\,
 \Theta_{c_0,s}(\Lambda),
 \label{eq:Utilde-tail-app}\\
 \Theta_{c_0,s}(\Lambda)
 &:=
 J_{c_0}\!\left(
 s+\tau_{c_0}(\Lambda/2)
 \right)^{1/2}
 +
 J_{c_0}\!\left(
 s+\tau_{c_0}(\Lambda/2)
 \right),
\end{align}
and the same bound holds for $\widetilde U_{n,\infty}(s)$.
Because $\Theta_{c_0,s}(\Lambda)\to0$, self-adjointness and
$I=\mathsf P_\Lambda+\mathsf Q_\Lambda$ give
\begin{align}
 \|\widetilde U_{n,c}-\widetilde U_{n,\infty}\|
 \leq{}&
 \left\|
 \mathsf P_\Lambda
 (\widetilde U_{n,c}-\widetilde U_{n,\infty})
 \mathsf P_\Lambda
 \right\|
 \nonumber\\
 &+
 2\|\mathsf Q_\Lambda\widetilde U_{n,c}\|
 +
 2\|\mathsf Q_\Lambda\widetilde U_{n,\infty}\|.
\end{align}
First taking $c\to\infty$ at fixed $\Lambda$ and then
$\Lambda\to\infty$ proves Eq.~\eqref{eq:Utilde-conv-app}.
\hfill$\square$

The point of this lemma is that the finite part need not become small in
the nonrelativistic limit; it must converge to the corresponding
Lieb--Liniger trace-space contribution.

\subsubsection*{Weighted contact map}

Define
\begin{equation}
 X_{n,c}(s)
 =
 K_{n,c}(-s)^{-1/2}T_{n,c}(s),
 \qquad
 X_{n,\infty}(s)
 =
 K_{n,\infty}(-s)^{-1/2}T_{n,\infty}(s).
 \label{eq:X-def-app}
\end{equation}
Then
\begin{align}
 X_{n,c}-X_{n,\infty}
 ={}&
 [K_{n,c}^{-1/2}-K_{n,\infty}^{-1/2}]T_{n,c}
 \nonumber\\
 &+
 K_{n,\infty}^{-1/2}(T_{n,c}-T_{n,\infty}),
 \label{eq:X-difference-app}
\end{align}
where the common dependence on $-s$ has been suppressed on the right.
The second term tends to zero by
Eq.~\eqref{eq:T-norm-conv-app} and the uniform inverse-square-root
bound.

For the first term set
\[
 F_c=K_{n,c}(-s)^{-1/2}-K_{n,\infty}(-s)^{-1/2}.
\]
Since the $K$'s are multiplication operators, they commute with
$\mathsf P_\Lambda$ and $\mathsf Q_\Lambda$.  Thus
\[
 F_cT_{n,c}
 =
 F_c\mathsf P_\Lambda T_{n,c}
 +
 F_c\mathsf Q_\Lambda T_{n,c}.
\]
On the bounded-momentum region,
Eq.~\eqref{eq:kappa-local-app} and the common positive lower bound imply
\begin{equation}
 \left\|
 \mathsf P_\Lambda
 [K_{n,c}^{-1/2}-K_{n,\infty}^{-1/2}]
 \mathsf P_\Lambda
 \right\|
 \longrightarrow0
\end{equation}
for every fixed $\Lambda$.  On the complementary region,
\begin{equation}
 \|F_c\mathsf Q_\Lambda T_{n,c}\|
 \leq
 2\kappa_s^{-1/2}
 \|\mathsf Q_\Lambda T_{n,c}\|,
\end{equation}
which is uniformly small for large $\Lambda$ by
Eq.~\eqref{eq:QT-tight-app}.  Hence
\begin{equation}
 \|X_{n,c}(s)-X_{n,\infty}(s)\|
 \longrightarrow0.
 \label{eq:X-conv-app}
\end{equation}

\subsection{Interaction correction and norm-resolvent convergence}
\label{app:NR-final}

Set
\begin{equation}
 A_{n,c}(s)=I-\widetilde U_{n,c}(s),
 \qquad
 A_{n,\infty}(s)=I-\widetilde U_{n,\infty}(s).
\end{equation}
Then
\begin{equation}
 \Phi_{n,c}(-s)
 =
 K_{n,c}(-s)^{1/2}
 A_{n,c}(s)
 K_{n,c}(-s)^{1/2},
\end{equation}
and
\begin{equation}
 \Phi_n^{\rm LL}(-s)
 =
 K_{n,\infty}(-s)^{1/2}
 A_{n,\infty}(s)
 K_{n,\infty}(-s)^{1/2}.
\end{equation}

Choose $s>\varepsilon_B$ sufficiently large that
$-s\in\rho(H_n^{\rm LL})$.  Then
$\Phi_n^{\rm LL}(-s)$ is invertible, and because
$K_{n,\infty}(-s)$ is bounded away from zero,
$A_{n,\infty}(s)$ is invertible.  By
Eq.~\eqref{eq:Utilde-conv-app},
\begin{equation}
 \|A_{n,c}(s)-A_{n,\infty}(s)\|
 \longrightarrow0.
\end{equation}
Therefore, for sufficiently large $c$,
\[
 \left\|
 A_{n,\infty}(s)^{-1}
 [A_{n,c}(s)-A_{n,\infty}(s)]
 \right\|<\frac12.
\]
The Neumann-series argument gives invertibility of $A_{n,c}(s)$ and the
uniform bound
\begin{equation}
 \sup_{c\geq c_1}\|A_{n,c}(s)^{-1}\|
 \leq
 2\|A_{n,\infty}(s)^{-1}\|.
 \label{eq:Ainv-bound-app}
\end{equation}
The inverse identity then implies
\begin{equation}
 A_{n,c}(s)^{-1}
 \xrightarrow[c\to\infty]{\|\cdot\|}
 A_{n,\infty}(s)^{-1}.
 \label{eq:Ainv-conv-app}
\end{equation}

The semirelativistic interaction correction is
\begin{align}
 T_{n,c}(s)^\dagger
 \Phi_{n,c}(-s)^{-1}
 T_{n,c}(s)
 =
 X_{n,c}(s)^\dagger
 A_{n,c}(s)^{-1}
 X_{n,c}(s),
\end{align}
whereas the Lieb--Liniger correction is
\begin{align}
 T_{n,\infty}(s)^\dagger
 [\Phi_n^{\rm LL}(-s)]^{-1}
 T_{n,\infty}(s)
 =
 X_{n,\infty}(s)^\dagger
 A_{n,\infty}(s)^{-1}
 X_{n,\infty}(s).
\end{align}
Using Eq.~\eqref{eq:X-conv-app},
Eq.~\eqref{eq:Ainv-conv-app}, and the uniform inverse bound, one obtains
\begin{align}
 \Big\|
 &T_{n,c}(s)^\dagger
 \Phi_{n,c}(-s)^{-1}
 T_{n,c}(s)
 \nonumber\\
 &-
 T_{n,\infty}(s)^\dagger
 [\Phi_n^{\rm LL}(-s)]^{-1}
 T_{n,\infty}(s)
 \Big\|
 \longrightarrow0.
 \label{eq:int-corr-conv-app}
\end{align}

Finally,
\begin{align}
 &(H_{n,c}-nmc^2+s)^{-1}
 \nonumber\\
 &\quad=
 R_{0,n}^{(c)}(-s)
 +
 T_{n,c}(s)^\dagger
 \Phi_{n,c}(-s)^{-1}
 T_{n,c}(s),
\end{align}
while
\begin{align}
 &(H_n^{\rm LL}+s)^{-1}
 \nonumber\\
 &\quad=
 R_{0,n}^{\rm NR}(-s)
 +
 T_{n,\infty}(s)^\dagger
 [\Phi_n^{\rm LL}(-s)]^{-1}
 T_{n,\infty}(s).
\end{align}
Combining the free-resolvent convergence with
Eq.~\eqref{eq:int-corr-conv-app} gives
\begin{equation}
 \lim_{c\to\infty}
 \left\|
 (H_{n,c}-nmc^2+s)^{-1}
 -
 (H_n^{\rm LL}+s)^{-1}
 \right\|
 =
 0.
 \label{eq:final-NR-norm-app}
\end{equation}

Restoring $\hbar$, the limiting operator is the self-adjoint
Lieb--Liniger Hamiltonian associated with
Eq.~\eqref{eq:LL-form-app}, formally
\begin{equation}
 H_n^{\rm LL}
 =
 \sum_{j=1}^{n}\frac{\hat p_j^{\,2}}{2m}
 -
 g_{\rm LL}\sum_{i<j}\delta(x_i-x_j),
 \qquad
 \hat p_j=-i\hbar\frac{\partial}{\partial x_j},
\end{equation}
with
\begin{equation}
 g_{\rm LL}
 =
 2\hbar\sqrt{\frac{\varepsilon_B}{m}},
 \qquad
 E_B(c)=2mc^2-\varepsilon_B.
\end{equation}
Thus, for every fixed particle number $n$,
\begin{equation}
 H_{n,c}-nmc^2
 \xrightarrow[c\to\infty]{\rm norm\ resolvent}
 H_n^{\rm LL}.
 \label{eq:final-NR-H-app}
\end{equation}
The proof establishes convergence at one sufficiently negative real
spectral point.  The resolvent identity extends it to the common
resolvent set.  The result is sectorwise: constants may depend on $n$,
and no uniform norm-resolvent convergence of the full Fock-space direct
sum is asserted.

\end{document}